\documentclass[]{imsart}
\arxiv{stats.ME/1406.7674}
\usepackage{amssymb}
\usepackage{amsmath}
\usepackage{graphicx}
\usepackage{subfigure}
\usepackage{breakcites}
\usepackage[authoryear]{natbib}
\usepackage{epsfig,amsfonts,amsmath,enumerate,graphicx,amssymb,bm,color,amsthm}
\usepackage{hyperref}
\usepackage{algorithm}
\usepackage[]{algorithmic}
\usepackage{xspace}

\newtheorem{theorem}{Theorem}

\newtheorem{corollary}{Corollary}

\newtheorem{remark}{Remark}

\begin{document}

\begin{frontmatter}
\title{Bayesian modelling of skewness and kurtosis with two-piece scale and shape distributions}
\runtitle{Bayesian modelling of skewness and kurtosis}

\begin{aug}

\author{F. J. Rubio\ead[label=e1]{francisco.rubio@warwick.ac.uk}}
\and
\author{M. F. J. Steel\ead[label=e2]{M.F.Steel@stats.warwick.ac.uk}}

\address{University of Warwick, Department of Statistics, Coventry, CV4 7AL, UK.,\\\printead{e1,e2}}

\runauthor{F. J. Rubio and M. F. J. Steel}
\end{aug}

\maketitle
\begin{abstract}
We formalise and generalise the definition of the family of univariate double two--piece distributions, obtained by using a density--based transformation of unimodal symmetric continuous distributions with a shape parameter. The resulting distributions contain five interpretable parameters that control the mode, as well as the scale and shape in each direction. Four-parameter subfamilies of this class of distributions that capture different types of asymmetry are discussed. We propose interpretable scale and location-invariant benchmark priors and derive conditions for the propriety of the corresponding posterior distribution. The prior structures used allow for meaningful comparisons through Bayes factors within flexible families of distributions. These distributions are applied to data from finance, internet traffic and medicine, comparing them with appropriate competitors.
\end{abstract}



\tableofcontents
\end{frontmatter}

\section{Introduction}\label{Introduction}

We present a generalisation of the two-piece transformation defined on the family of unimodal, continuous and symmetric univariate distributions that contain a shape parameter. This generalisation consists of using different scale and shape parameters either side of the mode. We call this the ``Double two-piece'' (DTP) construction. The resulting distributions contain five interpretable parameters that control the mode and the scale and shape in each direction. This transformation contains the original two-piece transformation as a subclass as well as a different class of transformations that only vary the shape of the distribution on each side of the mode. These two subclasses of distributions capture different types of asymmetry, recently denoted as ``main-body skewness'' and ``tail skewness'', respectively, by \cite{J14b}. {Although some particular members of the proposed DTP family have already been studied \citep{ZZ09,ZG10,ZG11}, we formalise this idea and extend it to a wider family of distributions, analysing the types of asymmetry that these distributions can capture. In addition, we propose and implement Bayesian methods for DTP distributions that allow us to meaningfully compare different distributions in these very flexible families through the use of Bayes factors. This directly sheds light on important  features of the data.} {As a byproduct, we propose a weakly informative prior elicitation strategy for the shape parameter of an arbitrary symmetric distribution. This strategy can be used, for example, to induce a proper prior for the degrees of freedom of the Student-$t$ distribution.}

In distribution theory, skewness and kurtosis are features of interest since they provide information about the shape of a distribution. Definitions and quantitative measures of these features have been widely discussed in the statistical literature (see e.g.~\citealp{Zwet64,GM84,CJ08}). Distributions containing parameters that control skewness and/or kurtosis are attractive since they can accommodate asymmetry and flexible tail behaviour. These types of flexible distributions are typically obtained by adding parameters to a known symmetric distribution through a parametric transformation. General representations of parametric transformations have been proposed in \cite{FS06} (probability integral transformations), \cite{LP10} (transformations of random variables) and \cite{J14a} (transformations of scale). Transformations that include a parameter that controls skewness are usually referred to as ``skewing mechanisms'' \citep{FS06,LP10} while those that add a kurtosis parameter have been called ``elongations'' \citep{FK04}, due to the effect produced on the shoulders and the tails of the distributions. Some examples of skewing mechanisms can be found in \cite{A85} and \cite{FS98a}. Examples of elongations can be found in \cite{T77}, \cite{H97}, \cite{FK04}, and \cite{KF06}. A third class of transformations consists of those that contain two parameters that are used for modelling skewness and kurtosis jointly. Some members of this class are the Johnson $\operatorname{S_U}$ family \citep{J49},  Tukey-type transformations such as the $g$-and-$h$ transformation and the LambertW transformation \citep{T77,G11}, and the sinh-arcsinh transformation \citep{JP09}. These sorts of transformations are typically, but not exclusively, applied to the normal distribution. Alternatively, distributions that can account for skewness and kurtosis can be obtained by introducing skewness into a symmetric distribution that already contains a shape parameter. Examples of distributions obtained by this method are skew-$t$ distributions \citep{H94,FS98a,AC03,RJP10}, and skew-Exponential power distributions \citep{A86,FOS95}. Other distributions containing shape and skewness parameters have been proposed in different contexts such as the generalized hyperbolic distribution \citep{BN77,AH06}, the skew--$t$ proposed in \cite{JF03}, and the $\alpha-$stable family of distributions. With the exception of the so called ``two--piece'' transformation \citep{FS98a,AGQ05}, the aforementioned transformations produce distributions with different shapes and/or different tail behaviour in each direction. Good surveys on families of flexible distributions can be found in \cite{J14b} and \cite{L14}. Finally, alternative approaches used to produce flexible models are semi-parametric models \citep{QSF09} or fully nonparametric models (\emph{e.g.}~kernel density estimators and Bayesian nonparametric density estimation). Some advantages of the models studied in this paper are the interpretability of the parameters and the ease of implementation in different contexts.

In Section \ref{Proposal}, we present the DTP construction and discuss some of its properties as well as two interesting subfamilies. We examine the nature of the asymmetry induced by these transformations and propose a useful reparameterisation. In Section \ref{Bayesian Inference} we present scale and location-invariant prior structures for the proposed models and derive conditions for the existence of the corresponding posterior distributions. Section \ref{Examples2} contains three examples using real data. The first two examples concern the fitting of internet traffic and financial data, and we show how DTP distributions can be used to better understand the asymmetry of these data. In a second type of application we study the use of DTP distributions to model the random effects in a Bayesian hierarchical model. We compare various flexible distributions in this context, using medical data. Proofs are provided in the Supplementary material. 


\section{Two-Piece Scale and Shape Transformations}\label{Proposal}

Let ${\mathcal F}$ be the family of continuous, unimodal, symmetric densities $\tilde f(\cdot;\mu,\sigma,\delta)$ with support on ${\mathbb R}$ and with mode and location parameter $\mu\in{\mathbb R}$, scale parameter $\sigma\in{\mathbb R}_+$, and shape parameter $\delta\in\Delta\subset{\mathbb R}$. A shape parameter is anything that is not a location or a scale parameter.

Denote $\tilde f(x;\mu,\sigma,\delta)=\dfrac{1}{\sigma}\tilde f\left(\dfrac{x-\mu}{\sigma};0,1,\delta\right)\equiv\dfrac{1}{\sigma}f\left(\dfrac{x-\mu}{\sigma};\delta\right)$. Distribution functions are denoted by the corresponding uppercase letters. We define the two-piece probability density function constructed of $f(x;\mu,\sigma_1,\delta_1)$ truncated to $(-\infty,\mu)$ and $f(x;\mu,\sigma_2,\delta_2)$ truncated to $[\mu,\infty)$:
\begin{equation}\label{dtppdf}
s(x;\mu,\sigma_1,\sigma_2,\delta_1,\delta_2) = \dfrac{2\varepsilon }{\sigma_1}f\left(\dfrac{x-\mu}{\sigma_1};\delta_1\right)I(x<\mu) + \dfrac{2(1-\varepsilon) }{\sigma_2}f\left(\dfrac{x-\mu}{\sigma_2};\delta_2\right)I(x\geq\mu),
\end{equation}
\noindent where we achieve a continuous density function if we choose
\begin{eqnarray}\label{epsilon}
\varepsilon = \dfrac{\sigma_1 f(0;\delta_2)}{\sigma_1 f(0;\delta_2)+\sigma_2 f(0;\delta_1)}.
\end{eqnarray}
We denote the family defined by (\ref{dtppdf}) and (\ref{epsilon}) as the Double Two-Piece (DTP) family of distributions. The corresponding cumulative distribution function is then given by
\begin{eqnarray}\label{dtpcdf}
S(x;\mu,\sigma_1,\sigma_2,\delta_1,\delta_2) &=& 2\varepsilon F\left(\dfrac{x-\mu}{\sigma_1};\delta_1\right)I(x<\mu) \notag\\
&+& \left\{\varepsilon +(1-\varepsilon)\left[2F\left(\dfrac{x-\mu}{\sigma_2};\delta_2\right)-1\right]\right\}I(x\geq\mu).
\end{eqnarray}

{The quantile function can be obtained by inverting (\ref{dtpcdf}).} By construction, the density (\ref{dtppdf}) is continuous, unimodal with mode at $\mu$, and the amount of mass to the left of its mode is given by $S(\mu;\mu,\sigma_1,\sigma_2,\delta_1,\delta_2)=\varepsilon$. This transformation preserves the ease of use of the original distribution $f$ and allows $s$ to have different shapes in each direction, dictated by  $\delta_1$ and $\delta_2$. In addition, by varying the ratio $\sigma_1/\sigma_2$, we control the allocation of mass on either side of the mode.

\begin{figure}[h!]
\begin{center}
\begin{tabular}{c c}
\psfig{figure=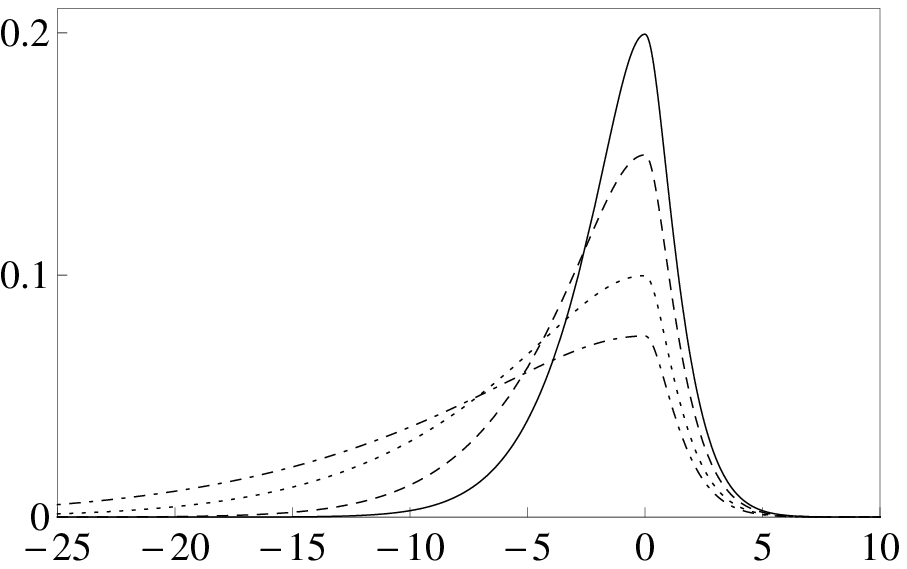,  height=4cm}&
\psfig{figure=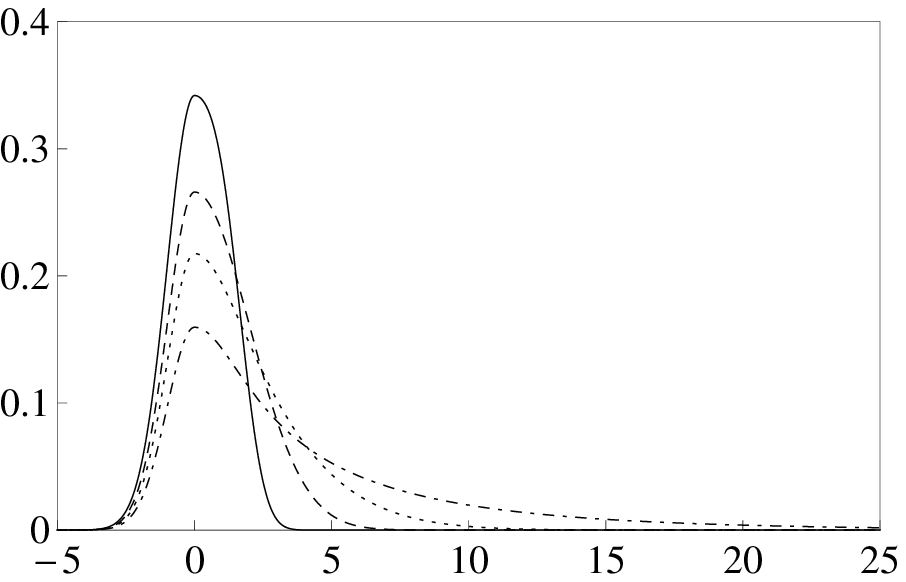,  height=4cm}\\
(a) & (b) 
\end{tabular}
\end{center}
\caption{\small DTP sinh-arcsinh (DTP SAS) distribution with $\mu=0$ and: (a) $\sigma_1=2,3,5,7$, $\sigma_2=1, \delta_1=\delta_2=0.75$; 
(b) $\sigma_1=1$, $\sigma_2=2$, $\delta_1=1$, $\delta_2=1.5,1,0.75,0.5$.}
\label{fig:shapes}
\end{figure}

The family ${\mathcal F}$, on which the proposed transformation is defined, can be chosen to be, for example, the symmetric Johnson-$\operatorname{S_U}$ distribution \citep{J49}, the symmetric sinh-arcsinh distribution \citep{JP09}, or the family of scale mixtures of normals, for which the density $f$ with shape parameter $\delta$ can be written as
$f(x_j; \delta) = \int_0^{\infty} \tau_j^{1/2}\phi(\tau_j^{1/2} x_j)d P_{\tau_j\vert\delta}$
for the observation $x_j$, where $\phi$ is the standard normal density and $P_{\tau_j\vert\delta}$ is a mixing distribution on ${\mathbb R}_+$. This is a broad class of distributions that includes, {\it i.a.}~the Student-$t$ distribution, the symmetric $\alpha-$stable distribution, the exponential power distribution {($1\leq\delta\leq2$)}, the symmetric hyperbolic distribution \citep{BN77}, and the symmetric $\alpha-$stable family (see \citealp{FS00} for a more complete overview). Here we also introduce the case where the mixing distribution is a Birnbaum-Saunders$(\delta,\delta)$ distribution, leading to what we call the SMN-BS distribution.
Expressions for the density of the SMN-BS and some other less common distributions are presented in the Appendix.
The shape parameter, $\delta>0$, in all these models can be interpreted as a kurtosis parameter. Figure $\ref{fig:shapes}$ illustrates the variety of shapes that we can obtain by applying the DTP transformation in (\ref{dtppdf}) to the symmetric sinh-arcsinh distribution. 

The DTP transformation  preserves the existence of moments, if and only if they exist for both $\delta_1$ and $\delta_2$, since
\begin{equation*}
\int_{\mathbb R} x^r s(x;\mu,\sigma_1,\sigma_2,\delta_1,\delta_2)dx = 2\varepsilon\int_{-\infty}^{\mu} x^r \tilde f(x;\mu,\sigma_1,\delta_1)dx
 + 2(1-\varepsilon)\int_{\mu}^{\infty} x^r \tilde f(x;\mu,\sigma_2,\delta_2)dx.
\end{equation*}
For example, if $f$ in $(\ref{dtppdf})$ is the Student-$t$ density with $\delta$ degrees of freedom, then the $r$th moment of $s$ exists if and only if both $\delta_1,\delta_2>r$.

A random variable with density (\ref{dtppdf}) can be decomposed as a variable that takes values distributed according to the density $2f(x;\mu,\sigma_1,\delta_1)I(x< \mu)$ with probability $\varepsilon$, while taking values distributed according to $2f(x;\mu,\sigma_2,\delta_2) I(x\geq\mu)$ with probability $1-\varepsilon$. Other distributions allow for more tangible stochastic representations, but these representations are typically based on untestable assumptions. For example, the distribution of the underlying selection mechanism in hidden truncation models (\citealp{AB02}), which include the skew-normal and skew-$t$ distributions of \cite{A85} and \cite{AC03} cannot be tested in practice. In addition, not all kinds of asymmetry are generated by hidden truncation and, in most contexts, the interest is not in modelling the underlying selection mechanism. \cite{J14b} argues that, although it is useful to have a tangible generating mechanism, we are often only interested in modelling skewness and kurtosis properly, so that the flexibility and inferential properties of the {final} model might be more important than the availability of an intuitive generating mechanism.

%


\subsection{Subfamilies with 4 Parameters}

\subsubsection*{Two-Piece Scale (TPSC) Distributions}

The DTP family of distributions naturally includes the original two--piece distribution by setting the condition $\delta_1=\delta_2=\delta$ in $(\ref{dtppdf})$, leading to
\begin{eqnarray}\label{tpsc}
s(x;\mu,\sigma_1,\sigma_2,\delta) = \dfrac{2}{\sigma_1+\sigma_2}\left[f\left(\dfrac{x-\mu}{\sigma_1};\delta\right)I(x<\mu) + f\left(\dfrac{x-\mu}{\sigma_2};\delta\right)I(x\geq\mu)\right].
\end{eqnarray}
The cases where $f(\cdot;\delta)$ is a Student-$t$ distribution or an exponential power distribution have already been analysed in some detail \citep{FOS95,FS98a}. 

\subsubsection*{Two-Piece Shape (TPSH) Distributions}

\begin{figure}[htp]
\begin{center}
\begin{tabular}{c c}
\psfig{figure=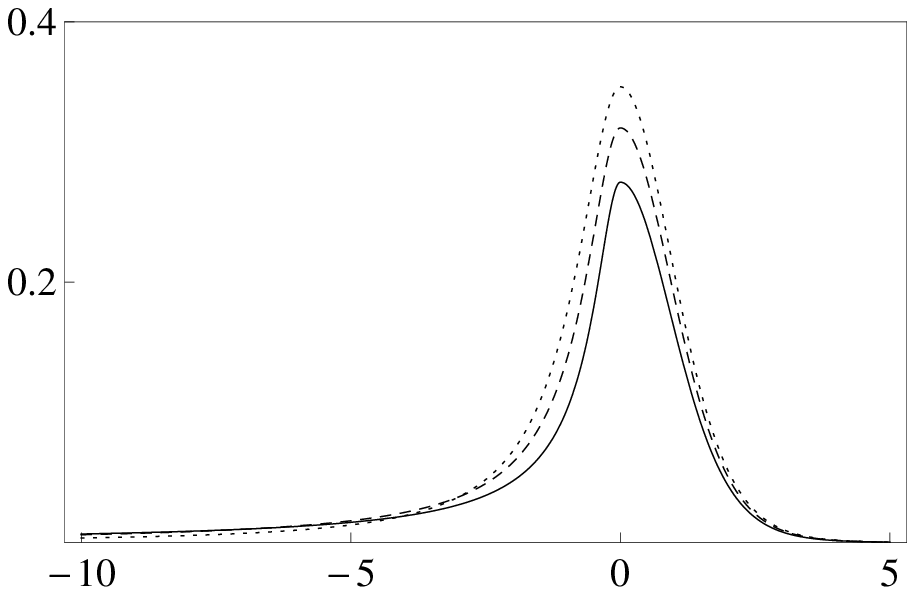,  height=4cm}  &
\psfig{figure=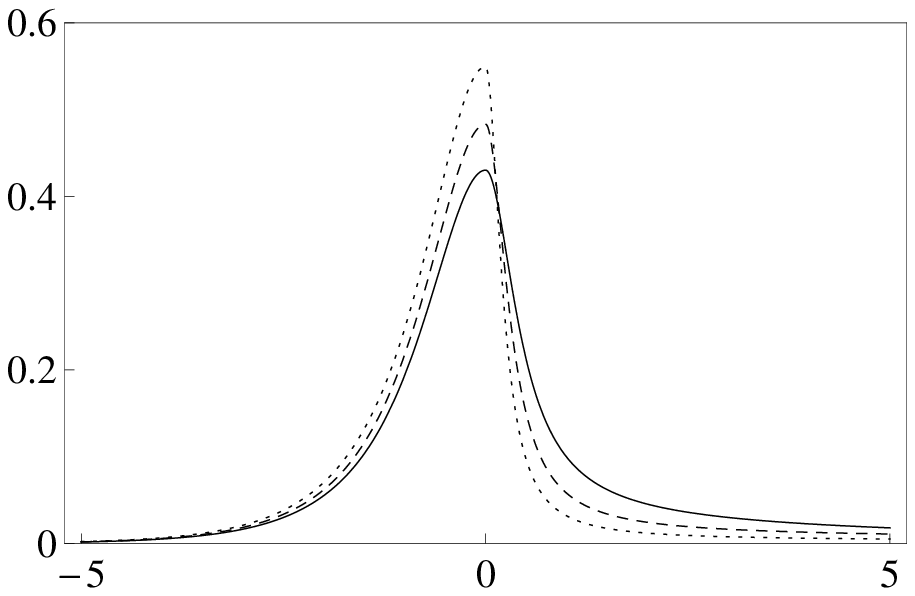,  height=4cm} \\
(a) & (b) \\
\end{tabular}
\end{center}
\caption{\small  TPSH densities with $(\mu,\sigma)=(0,1)$: (a) TPSH Student-$t$, $\delta_1=0.25,0.5,1$, $\delta_2=10$; (b) TPSH SMN-BS, $\delta_1=1$, $\delta_2=5,10,20$.}
\label{fig:dtshapes}
\end{figure}

An alternative subfamily can be obtained by fixing $\sigma_1=\sigma_2=\sigma$ in $(\ref{dtppdf})$, implying
\begin{eqnarray}\label{tpsh}
s(x;\mu,\sigma,\delta_1,\delta_2) &=& \dfrac{2\varepsilon }{\sigma}f\left(\dfrac{x-\mu}{\sigma};\delta_1\right)I(x<\mu)
 + \dfrac{2(1-\varepsilon) }{\sigma}f\left(\dfrac{x-\mu}{\sigma};\delta_2\right)I(x\geq\mu),
\end{eqnarray}
\noindent where $\varepsilon=\dfrac{f(0;\delta_2)}{f(0;\delta_1)+f(0;\delta_2)}$. 
This transformation produces distributions with different shape parameters in each direction. The variety of shapes obtained for different values of the parameters $(\delta_1,\delta_2)$ depends, of course, on the choice of the underlying symmetric model $f$. Note also that $\varepsilon$, the mass cumulated to the left of the mode, differs from $1/2$ whenever $f(0;\delta_1)\neq f(0;\delta_2)$. In the TPSH subclass skewness can only be introduced if the shape parameters differ in each direction. Other distributions with parameters that can control the tail behaviour in each direction have been proposed, for instance, in \cite{JF03}, \cite{AH06}, and \cite{JP09}. Figure \ref{fig:dtshapes} shows two examples of distributions obtained with the TPSH transformation. Interchanging $\delta_1$ and $\delta_2$ reflects the density function around the mode.

\subsection{Understanding the Skewing Mechanism Induced by the Proposed Transformations}\label{UnderstandSkewingMech}

In order to provide more insight into the family of DTP distributions, we analyse the TPSC and TPSH families of distributions separately. For this purpose we employ two measures of asymmetry defined for continuous unimodal distributions, the Critchley-Jones (CJ) functional asymmetry measure \citep{CJ08} and the Arnold-Groeneveld (AG) scalar measure of skewness \citep{AG95}. These measures of asymmetry are based on quantiles of the distributions, so they do not require the existence of moments such as the Pearson measure of skewness or the standardised third moment. Here we focus on the use of AG and CJ as measures of asymmetry due to their interpretability and the fact they are always well-defined. The CJ functional measures discrepancies between points located on each side of the mode $(x_L(p),x_R(p))$ of the density $g$ such that $g(x_L(p))=g(x_R(p))=pg(\mbox{mode})$, $p \in (0,1)$. It is defined as follows
\begin{eqnarray}\label{AsymFun}
\operatorname{CJ}(p)=\dfrac{x_R(p)-2\times\mbox{mode}+x_L(p)}{x_R(p)-x_L(p)}.
\end{eqnarray}
Note that this measure takes values in $(-1,1)$; negative values of $\operatorname{CJ}(p)$ indicate that the values $x_L(p)$ are further from the mode than the values $x_R(p)$. An analogous interpretation applies to positive values. The $\operatorname{AG}$ measure of skewness is defined as $1-2G(\mbox{mode})$, where $G$ is the distribution function associated with $g$. This measure also takes values in $(-1,1)$; negative values of $\operatorname{AG}$ are associated with left skewness and positive values correspond to right skewness. For the DTP family in $(\ref{dtppdf})$ these quantities are easy to calculate since $\operatorname{AG}=1-2\varepsilon$, and
\begin{eqnarray}\label{AFProp}
\operatorname{CJ}(p)=\dfrac{\sigma_2 f^{-1}_R(p f(0;\delta_2);\delta_2)+\sigma_1 f^{-1}_L(p f(0;\delta_1);\delta_1)}{\sigma_2 f^{-1}_R(p f(0;\delta_2);\delta_2)-\sigma_1 f^{-1}_L(p f(0;\delta_1);\delta_1)},
\end{eqnarray}
\noindent where $f^{-1}_L(\cdot;\delta)$ and $f^{-1}_R(\cdot;\delta)$ represent the negative and positive inverse of $f(\cdot;\delta)$, respectively. Note also that $\operatorname{CJ}(p)=\operatorname{AG}$ when $\delta_1=\delta_2$ for every $p\in(0,1)$. This means that for the TPSC family both measures coincide. {In general, the $\operatorname{AG}$ measure of skewness can be seen as an average of the asymmetry function $\operatorname{CJ}$ \citep{CJ08}}. In the TPSC family, asymmetry is produced by varying the scale parameters on each side of the mode. This simply reallocates the mass of the distribution while preserving the tail behaviour and the shape in each direction. Since the nature of the asymmetry induced by the TPSC transformation is intuitively rather straightforward and has been discussed in {\it e.g.}~\cite{FS98a}, we now focus on the study of TPSH transformations.

\begin{figure}[h!]
\begin{center}
\begin{tabular}{c c}
\psfig{figure=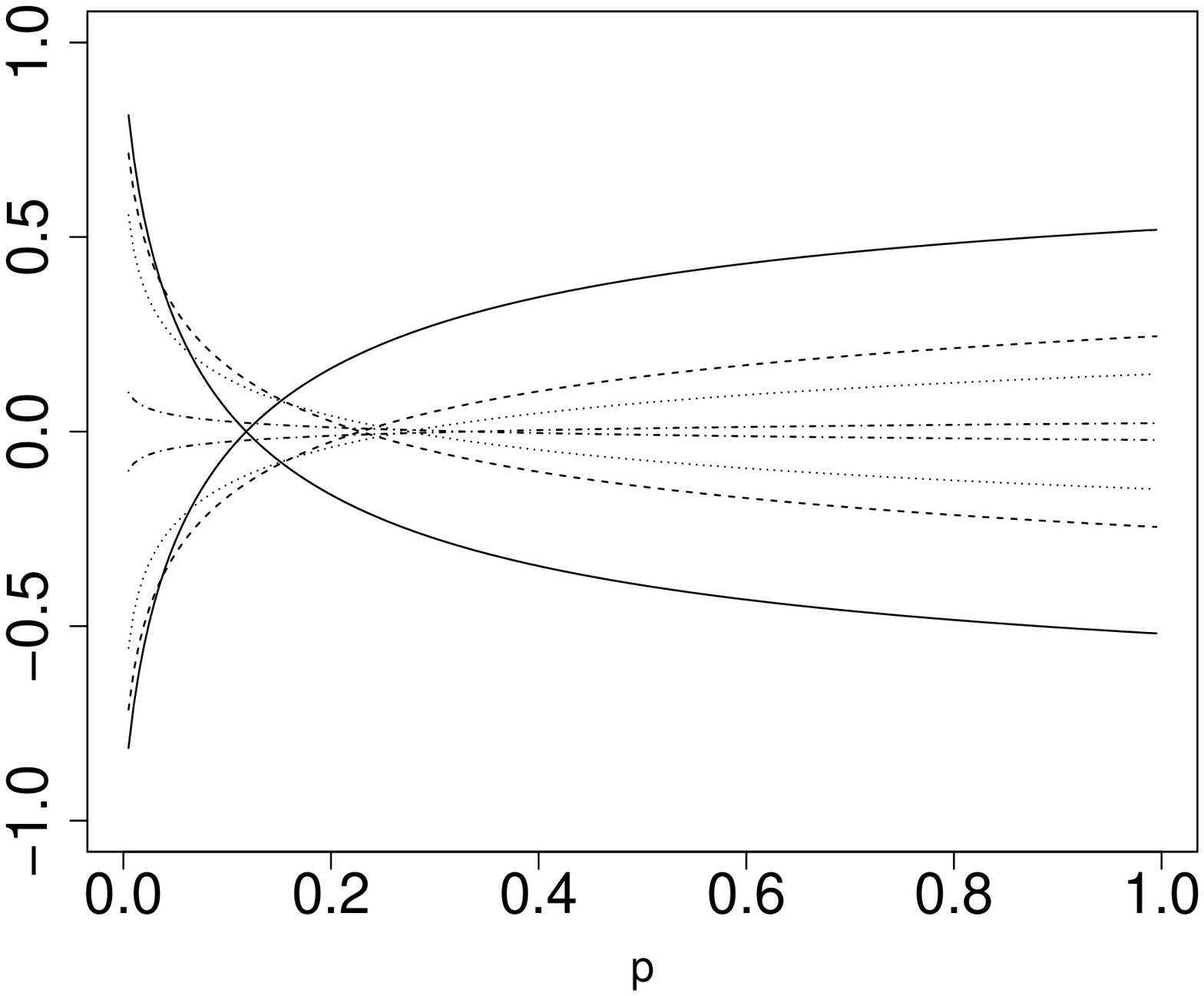,  height=4cm}  &
\psfig{figure=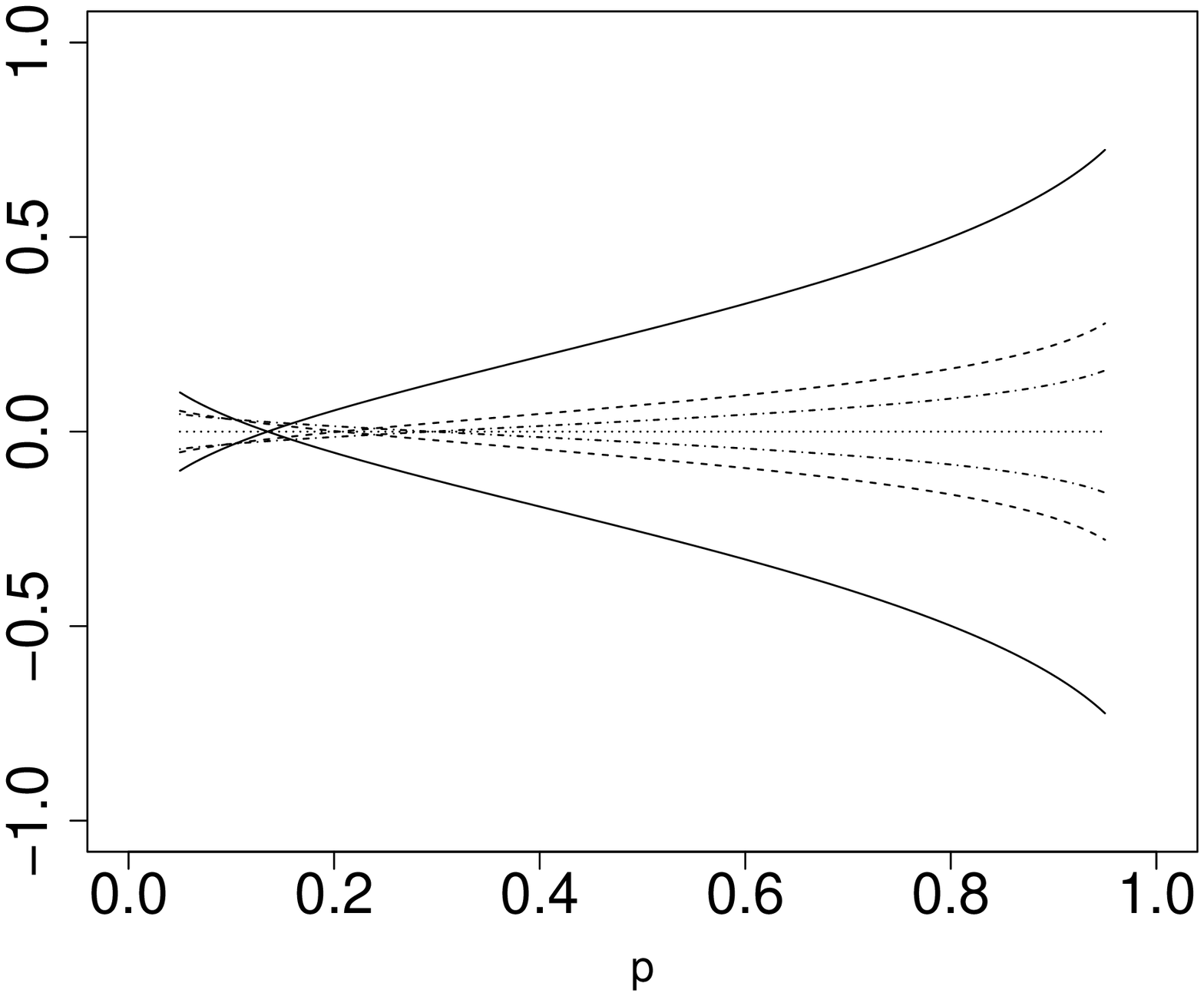,  height=4cm} \\
(a) & (b) \\
\psfig{figure=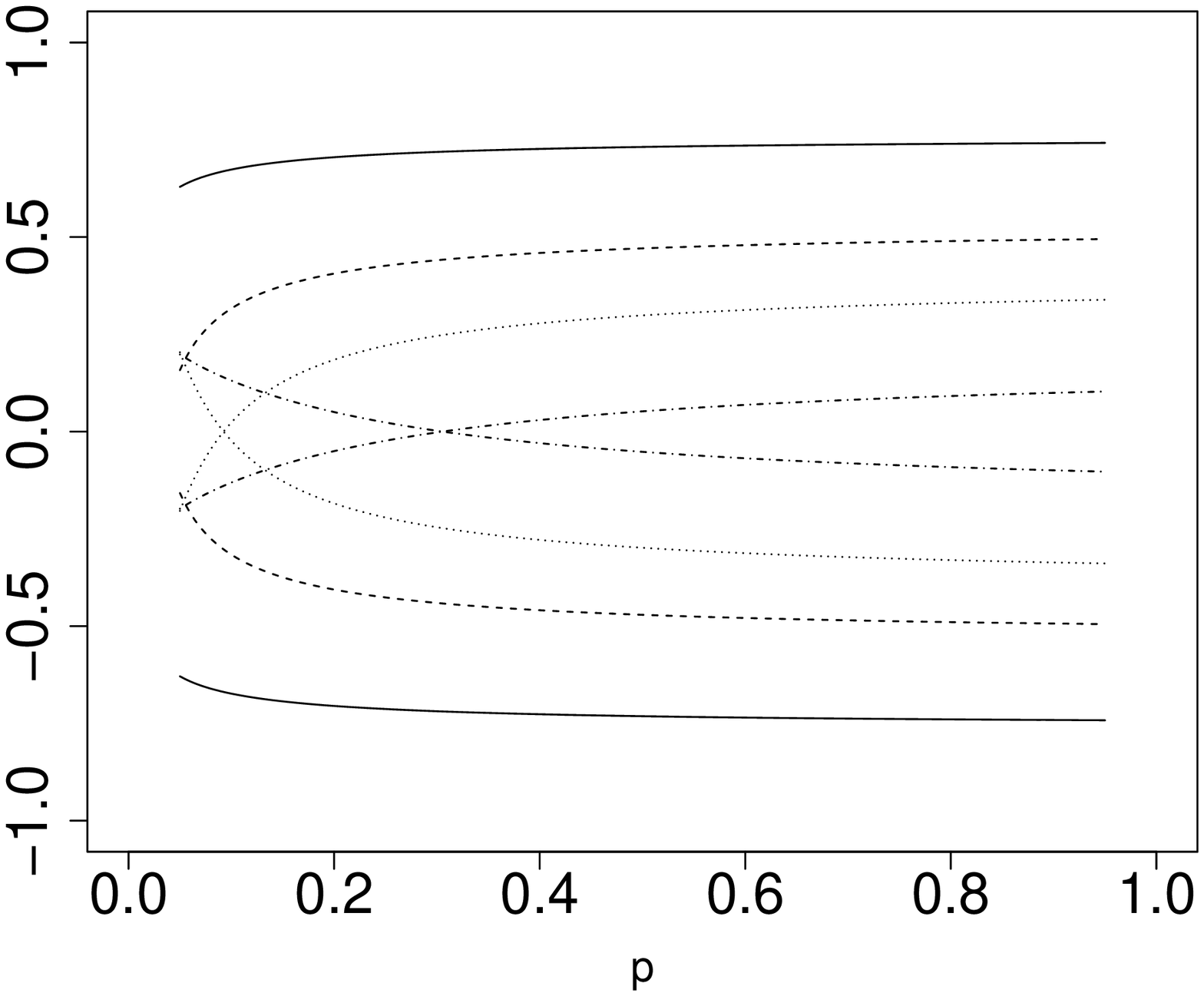,  height=4cm} &
\psfig{figure=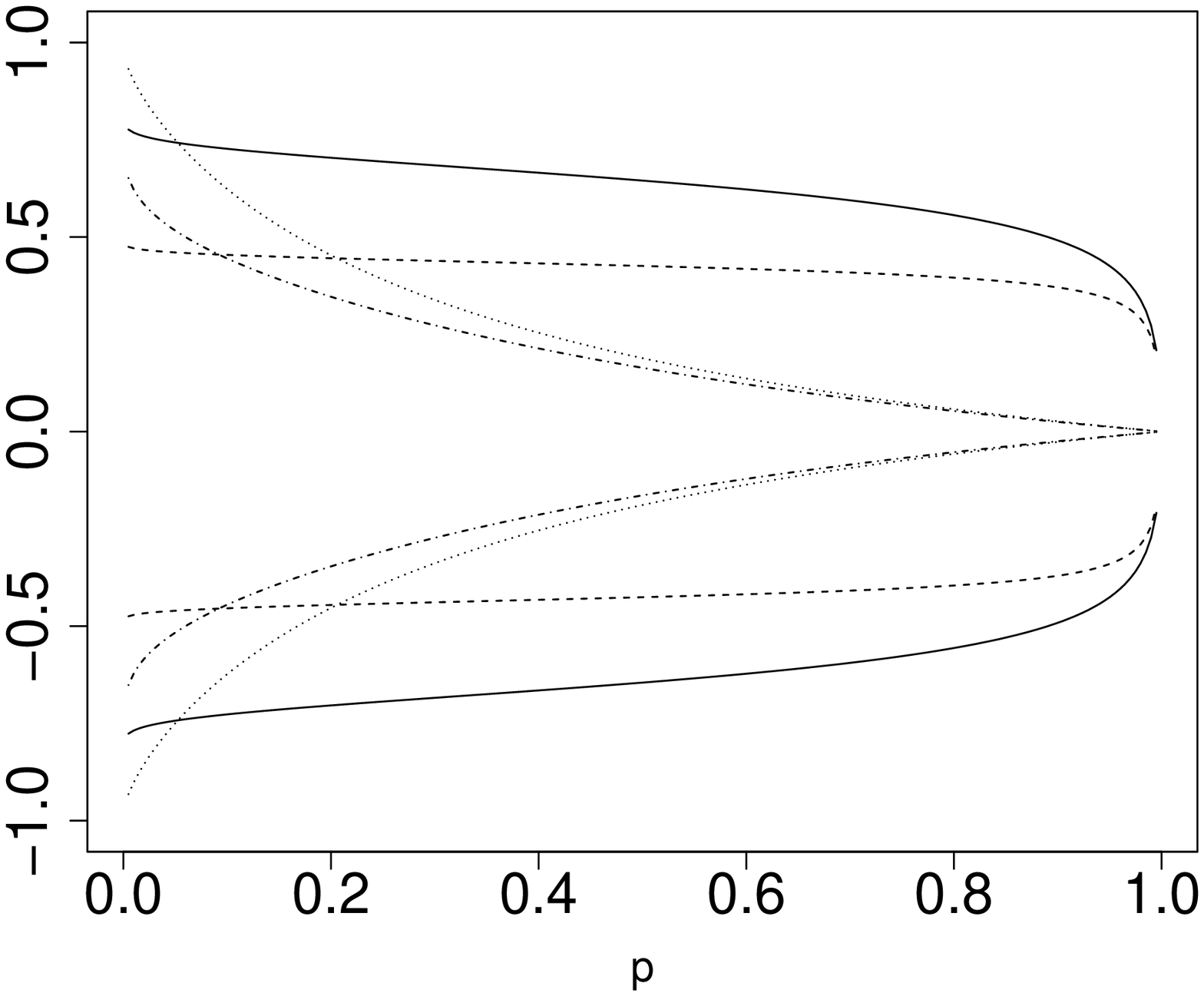,  height=4cm}\\
(c) & (d)
\end{tabular}
\end{center}
\caption{\small  Asymmetry functional $\operatorname{CJ}$ for: (a) TPSH Student $t$; (b) TPSH exponential power; (c) TPSH SMN-BS; (d) TPSH sinh-arcsinh distribution. Lines correspond to $\delta_1$ and $\delta_2$ as in Table \ref{table:AFPar} and those values reversed. }
\label{fig:AFSII}
\end{figure}

Figure \ref{fig:AFSII} shows some examples of (\ref{AFProp}) with distributions obtained using the TPSH transformation with parameters and AG as in Table \ref{table:AFPar}. Figures \ref{fig:AFSII}(a) and \ref{fig:AFSII}(b) show examples where $\operatorname{CJ}(p)$ changes sign in cases where AG is nonzero. This means that the relative distance of the points $(x_L(p),x_R(p))$ to the mode varies from the tails to the mode of the density as a consequence of the different shapes and clearly the TPSH transformation is quite different from the TPSC one (for which $\operatorname{CJ}$ is constant). Figure \ref{fig:AFSII}(c) corresponds to densities where $\operatorname{CJ}(p)$ changes sign for some combinations of the parameters $(\delta_1,\delta_2)$ while retaining the same sign for others. Finally, in Figure \ref{fig:AFSII}(d) $\operatorname{CJ}(p)$ retains the same sign for each $p$. Note that $\operatorname{CJ}$  for the SMN-BS distribution does not vary much with $p$, which means that TPSH and TPSC transformations are not that different. For the Student-$t$ and exponential power distributions (see Figures \ref{fig:AFSII}(a) and \ref{fig:AFSII}(b)) changing scale and shape parameters has very different consequences: skewness (as measured by $\operatorname{AG}$) is only induced for extremely low values of one of the shape parameters and the link between shape parameters and skewness (as measured by $\operatorname{CJ}(p)$) does not have a well-defined sign.

\begin{table}[h!]
\begin{center}
\begin{tabular}[h]{|c|c|c|c|c|c|c|c|c|c|c|c|}
\hline
\multicolumn{3}{|c|}{TPSH Student-$t$} & \multicolumn{3}{c|}{TPSH sinh-arcsinh} & \multicolumn{3}{c|}{TPSH SMN-BS} & \multicolumn{3}{c|}{TPSH exp. power}\\
\hline
$\delta_1$ & $\delta_2$ & $\operatorname{AG}$ &  $\delta_1$ & $\delta_2$ & $\operatorname{AG}$ &   $\delta_1$ & $\delta_2$  & $\operatorname{AG}$ &   $\delta_1$ & $\delta_2$  & $\operatorname{AG}$ \\
\hline
 1/10   & 10   & -0.45 &  5   & 1   & 2/3   &  1  &  50  &  -0.44  & 1   & 2   & 0.11  \\
\hline
1/2     & 10   & -0.18 &  5   & 2   & 0.43  &  1  &  10  & -0.09  & 1.5 & 2   & 0.03  \\
\hline
1       & 10   & -0.1  & 1    & 1/4 & 3/5   &  1  &  5  &  0.03  & 2   & 2   &  0 \\
\hline
5       & 10   & -0.01 &  1   & 1/2 & 1/3   &  2  &  1  &  -0.07  & 2.5 & 2   & -0.01  \\
\hline
\end{tabular}
\caption{\small Parameters used to obtain the functionals in Figure \ref{fig:AFSII}.}\label{table:AFPar}
\end{center}
\end{table}

\subsection{Reparameterisations}\label{reparameterisations}

For the TPSC family (\ref{tpsc}), \cite{AGQ05} propose the reparameterisation $(\mu,\sigma_1,\sigma_2,\delta)\leftrightarrow (\mu,\sigma,\gamma,\delta)$ using the transformation $\sigma_1=\sigma b(\gamma)$, $\sigma_2=\sigma a(\gamma)$, where $\{a(\cdot),b(\cdot)\}$ are positive differentiable functions, $\gamma\in\Gamma\subset{\mathbb R}$, and the parameter space $\Gamma$ depends on the choice of $\{a(\cdot),b(\cdot)\}$. The most common choices for $a(\cdot)$ and $b(\cdot)$ correspond to the \emph{inverse scale factors} parameterisation $\{a(\gamma),b(\gamma)\}=\{\gamma,1/\gamma\}$, $\gamma \in {\mathbb R}_+$ \citep{FS98a}, and the $\epsilon-$\emph{skew} parameterisation $\{a(\gamma),b(\gamma)\}=\{1-\gamma,1+\gamma\}$, $\gamma \in (-1,1)$ \citep{MH00}. \cite{JA10} and \cite{RS13b} show that choosing $a(\gamma)+b(\gamma)$ to be constant induces orthogonality between $\sigma$ and $\gamma$. This reparameterisation is also appealing because  the scalar $\gamma$ can be interpreted as a skewness parameter since the CJ and AG measures of skewness depend only on this parameter. In particular, we obtain
\[\operatorname{AG}=\frac{a(\gamma)-b(\gamma)}{a(\gamma)+b(\gamma)}.
\]

Moreover, \cite{KF06} showed that the parameter $\gamma$ can also be interpreted as a skewness parameter in terms of the partial ordering proposed by \cite{Zwet64}. This reparameterisation can also be used in DTP distributions for inducing orthogonality between $\sigma$ and $\gamma$ {through parameterisations that satisfy $a(\gamma)+b(\gamma)=\text{constant}$.  Under this reparameterisation, density (\ref{dtppdf}) becomes
\begin{equation}\label{dtppdfrepar}
s(x;\mu,\sigma,\gamma,\delta_1,\delta_2) = \dfrac{2}{\sigma c(\gamma,\delta_1,\delta_2)}\Biggl[f(0;\delta_2)f\left(\dfrac{x-\mu}{\sigma b(\gamma)};\delta_1\right)I(x<\mu)
 + f(0;\delta_1)f\left(\dfrac{x-\mu}{\sigma a(\gamma)};\delta_2\right)I(x\geq \mu)\Biggr],
\end{equation}
where $c(\gamma,\delta_1,\delta_2)=b(\gamma)f(0;\delta_2)+a(\gamma)f(0;\delta_1)$. The interpretation of $\gamma$ in the wider DTP family is slightly different since the cumulation of mass (and thus AG) depends also on the shape parameters $(\delta_1,\delta_2)$. However, the parameter $\gamma$ does not modify the shape of $s$.

Using this reparameterisation we can obtain the ``generalized asymmetric Student-$t$ distribution'' proposed in \cite{ZG10} by taking $f$ to be a Student-$t$ density and $\{a(\gamma),b(\gamma)\}=\{\gamma,1-\gamma\}$, $\gamma\in(0,1)$. Under the same parameterisation, the ``generalized asymmetric exponential power distribution'' proposed in \cite{ZZ09} corresponds to an exponential power density for $f$.

For the TPSH family (\ref{tpsh}) there seems to be no obvious reparameterisation that induces parameter orthogonality between the shape parameters and the other parameters. However, we can employ the reparameterisation $\delta_1=\delta b^*(\zeta)$, $\delta_2=\delta a^*(\zeta)$, with $\{a^*(\cdot),b^*(\cdot)\}$ positive differentiable functions. This  helps to separate the roles of the shape parameters, since $\delta$ can be interpreted as in the underlying symmetric model, while $\zeta$ explains the difference between the shapes on either side of the mode. The latter follows by noting that $\delta_1/\delta_2 = b^*(\zeta)/a^*(\zeta)$. This reparameterisation can also be applied to the DTP family, leading to the following density
\begin{eqnarray}\label{dtppdfrepar2}
s(x;\mu,\sigma,\gamma,\delta,\zeta) &=& \dfrac{2}{\sigma c(\gamma,\delta,\zeta)}\Biggl[f(0;\delta a^*(\zeta))f\left(\dfrac{x-\mu}{\sigma b(\gamma)};\delta b^*(\zeta)\right)I(x<\mu)\notag\\
 &+& f(0;\delta b^*(\zeta))f\left(\dfrac{x-\mu}{\sigma a(\gamma)};\delta a^*(\zeta)\right)I(x\geq \mu)\Biggr],
\end{eqnarray}
where $c(\gamma,\delta,\zeta)=b(\gamma)f(0;\delta a^*(\zeta))+a(\gamma)f(0;\delta b^*(\zeta))$.
\section{Bayesian Inference}\label{Bayesian Inference}

\subsection{Improper priors and posterior propriety}\label{Improper Priors}

In this section we propose a class of ``benchmark'' priors for the models studied in Section \ref{Proposal} with the parameterisations in (\ref{dtppdfrepar}) or (\ref{dtppdfrepar2}). The proposed prior structure is inspired by the independence Jeffreys prior and the reference prior for the symmetric model, producing a scale and location-invariant prior.

The following result shows that the use of improper priors on the shape parameters of DTP models often leads to improper posteriors.

\begin{theorem}\label{ImproperPosterior}
Let ${\bf x}=(x_1,...,x_n)$ be an independent sample from $(\ref{dtppdfrepar})$ and consider the prior structure
\begin{eqnarray}\label{priorStructureDTP}
p(\mu,\sigma,\gamma,\delta_1,\delta_2) \propto p(\mu)p(\sigma)p(\gamma)p(\delta_1)p(\delta_2),
\end{eqnarray}
\noindent where $p(\delta_1)$ and/or $p(\delta_2)$ are improper priors.
\begin{enumerate}[(i)]

\item If $f(0;\delta)$ does not depend upon $\delta$, then the posterior is improper.

\item If $f(0;\delta)$ is bounded from above, then a necessary condition for posterior propriety is
\begin{eqnarray}\label{NecCondI}
\int_{\Delta} f(0;\delta_i)^n p(\delta_i) d\delta_i<\infty,\,\,\, i=1,2.
\end{eqnarray}

\item If $f(0;\delta)$ is a continuous and monotonic function of $\delta$, then for any $0\leq \inf_{\delta\in\Delta}f(0;\delta)<M < \sup_{\delta\in\Delta}f(0;\delta)$, a necessary condition for the propriety of the posterior is
\begin{eqnarray}\label{NecCondII}
\int_{\Delta} \dfrac{f(0;\delta_i)^n}{\left[f(0;\delta_i)+M\right]^n} p(\delta_i) d\delta_i<\infty,\,\,\, i=1,2.
\end{eqnarray}
\end{enumerate}

\end{theorem}

Clearly, conditions (\ref{NecCondI}) and (\ref{NecCondII}) are satisfied when $p(\delta_i)$ is proper for $i=1,2$, but they often do not hold under improper priors.
Thus, Theorem \ref{ImproperPosterior} provides a warning against the use of improper priors on the shape parameters of DTP models. For instance, (i), (ii) and (iii) imply, respectively, that the use of improper priors on the shape parameters $(\delta_1,\delta_2)$ of DTP exponential power (with the parameterisation in \citealp{ZZ09}; see \citealp{R14} for an example), DTP Student--$t$, and DTP sinh--arcsinh distributions leads to improper posteriors.

In the DTP model (\ref{dtppdfrepar}) the parameters $\gamma$ and $(\delta_1,\delta_2)$ control the difference in the scale and the shapes either side of the mode, respectively. So we adopt a product prior structure $p(\gamma)p(\delta_1,\delta_2)$, allowing for prior dependence between $\delta_1$ and $\delta_2$.  The following result provides conditions for the existence of the corresponding posterior distribution when $f$ is a scale mixture of normals. The case where the sample contains repeated observations is covered as well.

\begin{theorem}\label{DTPproper} 
Let ${\bf x}=(x_1,...,x_n)$ be an independent sample from $(\ref{dtppdfrepar})$. Let $f$ be a scale mixture of normals and consider the prior structure
\begin{eqnarray}\label{priorDTP}
p(\mu,\sigma,\gamma,\delta_1,\delta_2) \propto \dfrac{1}{\sigma}p(\gamma)p(\delta_1,\delta_2),
\end{eqnarray}
\noindent where $p(\gamma)$ and $p(\delta_1,\delta_2)$ are proper.
\begin{enumerate}[(i)]

\item The posterior distribution of $(\mu,\sigma,\gamma,\delta_1,\delta_2)$ is proper if $n\geq 2$ and all the observations are different.

\item If ${\bf x}$ contains repeated observations, let $k$ be the largest number of observations with the same value in  ${\bf x}$ and $1<k<n$, then the posterior of $(\mu,\sigma,\gamma,\delta_1,\delta_2)$ is proper if and only if the mixing distribution of $f$ satisfies for $i=1,2$ and $j$ the observation index
\begin{eqnarray}\label{condmix}
\int_{0<\tau_1\leq\dots\leq\tau_{n}<\infty} \tau_{n-k}^{-(n-2)/2}\prod_{j\neq n-k,n}\tau_j ^{1/2} d P_{(\tau_1,\dots,\tau_{n}|\delta_i)} d\delta_i<\infty.
\end{eqnarray}

In the case of a two-piece Student-$t$ sampling model, $(\ref{condmix})$ is equivalent to
\begin{eqnarray}\label{condt}
\int_{(k-1)/(n-k)}^{(k-1)/(n-k)+\xi}\dfrac{p(\delta_i)}{(n-k)\delta_i - (k-1)}d\delta_i<\infty \,\,\, \text{and}\,\, \int_{0}^{(k-1)/(n-k)}p(\delta_i)d\delta_i=0,\,\,\,\,\,
\end{eqnarray}
\noindent for all $\xi>0$ and $i=1,2$.
\end{enumerate}

\end{theorem}

For the reparameterisation (\ref{dtppdfrepar2}), the parameters $(\gamma,\delta,\zeta)$ have separate roles: $\gamma$ controls the difference in the scale either side of the mode, $\delta$ represents the shape parameter of the underlying symmetric density, and $\zeta$ controls the difference in the shape either side of the mode. For this reason, it is reasonable to adopt an independent prior structure on these parameters. The following result provides conditions for the existence of the posterior distribution.

\begin{remark}\label{RemarkRepar}
Let ${\bf x}=(x_1,...,x_n)$ be an independent sample from $(\ref{dtppdfrepar2})$. Let $f$ be a scale mixture of normals and consider the prior structure
\begin{eqnarray}\label{priorDTP2}
p(\mu,\sigma,\gamma,\delta,\zeta) \propto \dfrac{1}{\sigma}p(\gamma)p(\delta)p(\zeta),
\end{eqnarray}
\noindent where $p(\gamma)$, $p(\delta)$, and $p(\zeta)$ are proper. The posterior distribution of $(\mu,\sigma,\gamma,\delta,\zeta)$ is proper if $n\geq 2$ and all the observations are different. If the sample contains repeated observations, we need to check that the induced prior on $(\delta_1,\delta_2)$, for the parameterisation (\ref{dtppdfrepar}), satisfies (\ref{condmix}).

\proof The results follows by a change of variable from $(\delta_1,\delta_2)$ to $(\delta,\zeta)$.
\end{remark}

As discussed in previous sections, the parameters of a distribution obtained through the TPSC transformation,  $(\mu,\sigma,\gamma,\delta)$, can be interpreted as location, scale, skewness and shape, respectively. For this reason we adopt the product prior structure $p(\mu,\sigma,\gamma,\delta)\propto \dfrac{1}{\sigma}p(\gamma)p(\delta)$ for this family. In TPSH models the shape parameters $(\delta_1,\delta_2)$ control the mass cumulated on each side of the mode as well as the shape. In addition, these parameters are not orthogonal in general. We therefore adopt the product prior structure $p(\mu,\sigma,\delta_1,\delta_2)\propto \dfrac{1}{\sigma}p(\delta_1,\delta_2)$ in this family, where $p(\delta_1,\delta_2)$ denotes a proper joint distribution  which allows for prior dependence between $\delta_1$ and $\delta_2$. Theorem \ref{DTPproper} covers the propriety of the posterior under these priors for TPSC and TPSH sampling models. For TPSH models with the parameterisation (\ref{dtppdfrepar2}), Remark \ref{RemarkRepar} provides conditions for the existence of the posterior distribution under the prior $p(\mu,\sigma,\delta,\zeta)\propto \dfrac{1}{\sigma}p(\delta)p(\zeta)$.

Another context of practical interest is when the sample consists of set observations. A set observation $S$ is simply defined as a set of positive probability under the sampling model, {\it i.e.}~${\mathbb P}[\text{Observing }S] >0$. In particular, this corresponds to any observation recorded with finite precision, as well as left, right and interval censoring. When the quantitative effect of censoring is not negligible, this must be formally taken into account. 
The following corollary provides conditions for the existence of the posterior from set observations with DTP sampling models.

\begin{corollary}\label{SetObs}
Let ${\bf x}=(S_1,...,S_n)$ be an independent sample of set observations from $(\ref{dtppdfrepar})$. Let $f$ be a scale mixture of normals and consider the prior structure (\ref{priorDTP}). Then, the posterior distribution of $(\mu,\sigma,\gamma,\delta_1,\delta_2)$ is proper if $n\geq 2$ and there exists a pair of sets, say $(S_i,S_j)$, such that
\begin{eqnarray}\label{condsets}
\inf_{x_i \in S_i, x_j \in S_j} \vert x_i - x_j \vert &>&0.
\end{eqnarray}
\end{corollary}

Thus, whenever each sample of set observations contains at least two intervals that do not overlap, the posterior distribution of $(\mu,\sigma,\gamma,\delta_1,\delta_2)$ is proper. This result also applies to the parameterisation (\ref{dtppdfrepar2}) with prior (\ref{priorDTP2}).

\subsection{Choice of the prior on $(\gamma,\delta,\zeta)$}\label{PriorShapes}

We now propose specific priors for the parameters $(\gamma,\delta,\zeta)$ in (\ref{priorDTP2}) for a general choice of $f$ in $(\ref{dtppdfrepar2})$, and its corresponding subfamilies. We employ the parameterisation $\{a(\gamma),b(\gamma)\}= \{1-\gamma,1+\gamma\}$, $\gamma\in(-1,1)$ (so that $\sigma$ and $\gamma$ are orthogonal), and  $\{a^*(\zeta),b^*(\zeta)\}= \{1-\zeta,1+\zeta\}$, $\zeta\in(-1,1)$. The shape parameter $\delta$ typically controls the peakedness and the heaviness of tails of the density function.  As mentioned earlier, the parameters $\gamma$ and $\zeta$ control the difference in scale and shape either side of $\mu$. This interpretability of the parameters facilitates the choice of hyperparameters. 
In particular, reasonable priors to reflect vague prior beliefs are that $\gamma\sim \text{Unif}(-1,1)$  and $\zeta\sim \text{Unif}(-1,1)$. The elicitation of the prior on the parameter $\delta$ is more delicate, given that this parameter has different interpretations for different models. However, in all the models of interest, $\delta$ can be interpreted as a kurtosis parameter. Therefore, in order to come up with a more general elicitation strategy we propose basing this choice on a prior for a bounded kurtosis measure, which is common to all models and is an injective function of $\delta$, say $\kappa=\kappa(\delta)$. The boundedness assumption on $\kappa$ allows us to assign a proper uniform prior on this quantity, while the injectivity is required for obtaining the induced prior on the parameter $\delta$ by inverting this function. See \cite{CJ08} for a good survey on kurtosis measures. 

We propose to adopt the scalar kurtosis measure $\kappa = 2\dfrac{f(\pi_R)}{f(\text{mode})}-1$ from \cite{CJ08}, where $\pi_R$ represents the positive mode of $-f^{\prime}$ (the inflection point). 
This measure $\kappa$ takes values in $K\subset(-1,1)$, assigning the value $\kappa=0.213$ to the normal distribution. Numerically, we have found that $\kappa$ is an injective function of $\delta$ for many distributions $f$, such as the Student-$t$, the symmetric sinh-arcsinh, the symmetric Johnson-$S_U$, the exponential power with $\delta>1$, the symmetric hyperbolic, the SMN-BS with $\delta<2.65$, and the Meixner distribution. Another appealing feature of this measure of kurtosis is that both the AG skewness measure and $\kappa$ can be interpreted as the average of certain functional measures of asymmetry and kurtosis using the same weight function (see \citealp{CJ08}).  Figure \ref{fig:TSASPriors} shows the priors for $\delta$ for the Student-$t$ and symmetric sinh-arcsinh distributions,  induced  by a uniform prior on the appropriate range for $\kappa$.  {The prior for $\delta$ in the Student-$t$ model is an alternative to the Jeffreys prior in \cite{Fonseca} and is quite close to the gamma-gamma prior of \cite{Juarez} with their parameter $d=1.2$. It is also a continuous alternative to the discrete objective prior proposed in \cite{VW14}.}


\begin{figure}[!h]
\begin{center}
\begin{tabular}{c c}
\psfig{figure=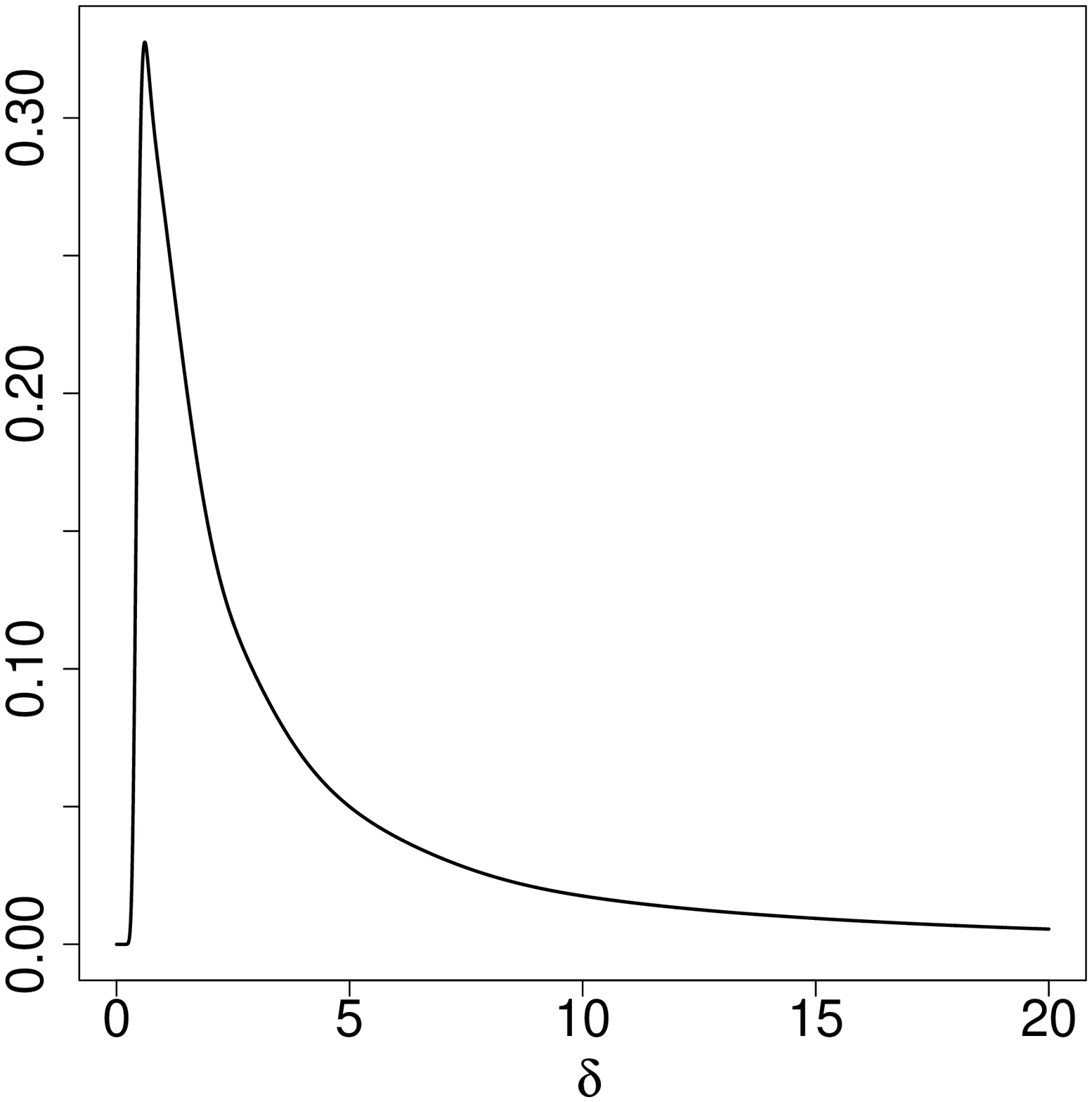,  height=4cm}  &
\psfig{figure=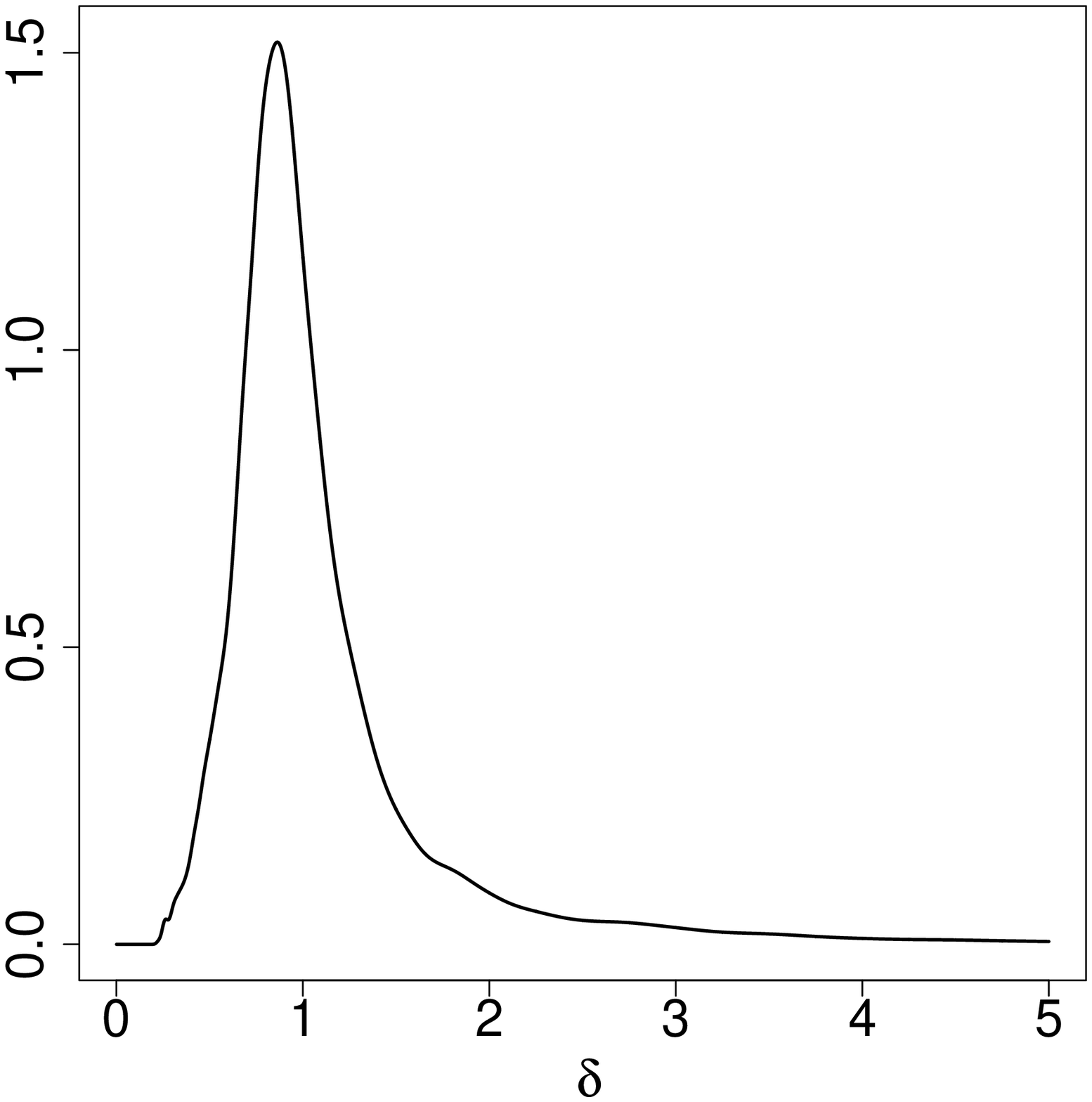,  height=4cm}\\
(a) & (b)
\end{tabular}
\end{center}
\caption{\small Priors for $\delta$: (a) Student-$t$ distribution; (b) Sinh-arcsinh distribution.}
\label{fig:TSASPriors}
\end{figure}

\subsection{Weakly informative proper priors}

We may prefer to use a ``vague'' proper prior which is not very influential on the posterior inference. 
In the previous section we provided weakly informative priors for the shape parameters $(\gamma,\delta,\zeta)$. We can combine that with independent vague proper priors on the location and scale parameters $(\mu,\sigma)$. For the location parameter we propose a uniform prior on an appropriate bounded interval $\mathcal{D}$, while for the scale parameter we employ a Half-Cauchy distribution with location $0$ and scale $s$ \citep{PS12}. Unfortunately, general choices for $\cal{D}$ and $s$ are not available, given that these values depend on the units of measurement. We recommend conducting sensitivity analyses with respect to $\mathcal{D}$ and $s$. Note that the structure of this prior resembles that of the improper benchmark priors discussed in the previous sections.

This prior structure is also useful for choices of $f$ that do not belong to the family of scale mixtures of normals and, consequently, the existence of the posterior under improper priors is not covered by the results in Subsection \ref{Improper Priors}.

\section{Applications}\label{Examples2}

We present three examples with real data to illustrate the use of DTP, TPSC and TPSH distributions. We adopt the $\epsilon-$skew parameterisation for DTP and TPSC models. In the first two examples, simulations of the posterior distributions are obtained using the $t$-walk algorithm \citep{CF10}. 
Given the hierarchical nature of the third example, we use the adaptive Metropolis within Gibbs sampler implemented in the R package `spBayes' \citep{FB07}. R codes used here and the R-package `DTP', which implements basic functions related to the proposed models, are available on request.

Model comparison within the DTP family is conducted via Bayes factors which are obtained using the Savage--Dickey ratio  for nested models, and through importance sampling when we compare non-nested choices for $f$. We also compare the DTP model and its submodels with other distributions used in the literature. For a fair model comparison, we include appropriate competitors in each example, matched to the features of the data. A meaningful Bayesian comparison with these other models would require the specification of priors for the parameters in these other distributions that are comparable (matched) to our models, and to compute Bayes factors we would need to use proper priors for all model-specific parameters. This would be a nontrivial undertaking and would risk diluting the main message of the paper. We choose instead to compare with these other classes of distributions through classical information criteria based on maximum likelihood estimates (MLE). We aim to show that the DTP families are flexible enough and then we can use formal Bayesian methods to select (or average) models within these families.

{Given that DTP, TPSC, and TPSH distributions capture different sorts of asymmetry, conducting model comparison between these distributions not only provides information about which model fits the data better but it also indicates what kind of asymmetry is favoured by the data. In addition, the DTP family provides important advantages in terms of interpretability of parameters (and, thus, prior elicitation) and inferential properties.}

\subsection{{Internet traffic data}}

In this example we analyse the teletraffic data set studied in \cite{R10}, which contains $n=3143$ observations, representing transferred bytes/sec within consecutive seconds. \cite{R10} propose the use of a Normal Laplace distribution to model these data after a logarithmic transformation. The Normal Laplace distribution is obtained as the convolution of a Normal distribution and a two--piece Laplace distribution with location $0$ and two parameters $(\alpha,\beta)$ that jointly control the scale and the skewness. The Normal Laplace distribution has tails heavier than those of the normal distribution \citep{RJ04}. We also use the sinh-arcsinh distribution of \cite{JP09}, indicated by $s_{JP}$ and the skew-$t$ of \cite{AC03}, denoted by $s_{AC}$ (see Appendix). Here, we explore the performance of the DTP sinh--arcsinh distribution (DTP SAS). 
This distribution allows for all moments to exist and accommodates both heavier and lighter tails than the normal distribution, which is a submodel of the DTP SAS ($\delta_1=\delta_2=1$, $\gamma=0$).
We use the priors of Subsection 3.3:
$\mu \sim \text{Unif}(0,25),
\sigma \sim \text{HalfCauchy}(0,s),
\gamma \sim \text{Unif}(-1,1),
\zeta \sim \text{Unif}(-1,1),$
where $s=1/5,1,5$ and for $\delta$ we adopt the prior in Figure \ref{fig:TSASPriors}. The results were not sensitive to the choice of $s$. Table \ref{table:Internet} shows the MLE and the classical model comparison criteria for all models considered. The DTP SAS results indicate that the right tail is much lighter than that of the normal distribution, a feature that cannot be captured by the Normal Laplace distribution used in \cite{R10}. In addition, there is strong evidence of ``main-body'' skewness, captured by different scales. Both features of the models are clearly important
for these data and the DTP SAS model is strongly favoured by AIC and BIC. Bayes factors within the DTP SAS family also strongly support the most complete model, versus the possible submodels (all of them are $<10^{-100}$). Posterior predictive densities shown in Figure \ref{fig:internetpred} illustrate how the DTP SAS model differs from the others in mode and tail behaviour (see the right panel).
\begin{table}[ht]
\begin{center}
\begin{tabular}[h]{|c|c|c|c|c|c|c|c|}
\hline
Model & $\widehat{\mu}$ & $\widehat{\sigma}$ & $\widehat{\gamma}$ & $\widehat{\delta}$  & $\widehat{\zeta}$ & AIC & BIC\\
\hline
DTP SAS & 11.15  & 13.82 & -0.98 & 12.95  & -0.95 & {\bf 5849.03} & {\bf 5879.29} \\
\hline
TPSC SAS & 11.80 & 0.85 & 0.14 & 1.26  & -- & 5884.95 &  5909.16 \\
\hline
TPSH SAS & 11.75 & 0.87 & -- & 1.30  & -0.08 & 5880.20 & 5904.41 \\
\hline
$s_{JP}$ & 11.78 & 0.84 & ($\widehat{\varepsilon}$) -0.16 & 1.25  & --  & 5886.84 & 5911.05 \\
\hline
Normal Laplace & 11.77 & 8.39 & ($\widehat{\alpha}$) 4.09 & ($\widehat{\beta}$) 0.56  & -- & 5922.73 & 5946.94 \\
\hline
$s_{AC}$ & 12.07 & 0.75 & ($\widehat{\lambda}$) -0.98 & 1057.40  & --  & 5919.52 & 5943.73 \\
\hline
\end{tabular}
\caption{\small Internet traffic data: Maximum likelihood estimates, AIC and BIC (best values in bold).}
\label{table:Internet}
\end{center}
\end{table}

\begin{figure}[!h]
\begin{center}
\begin{tabular}{c c}
\psfig{figure=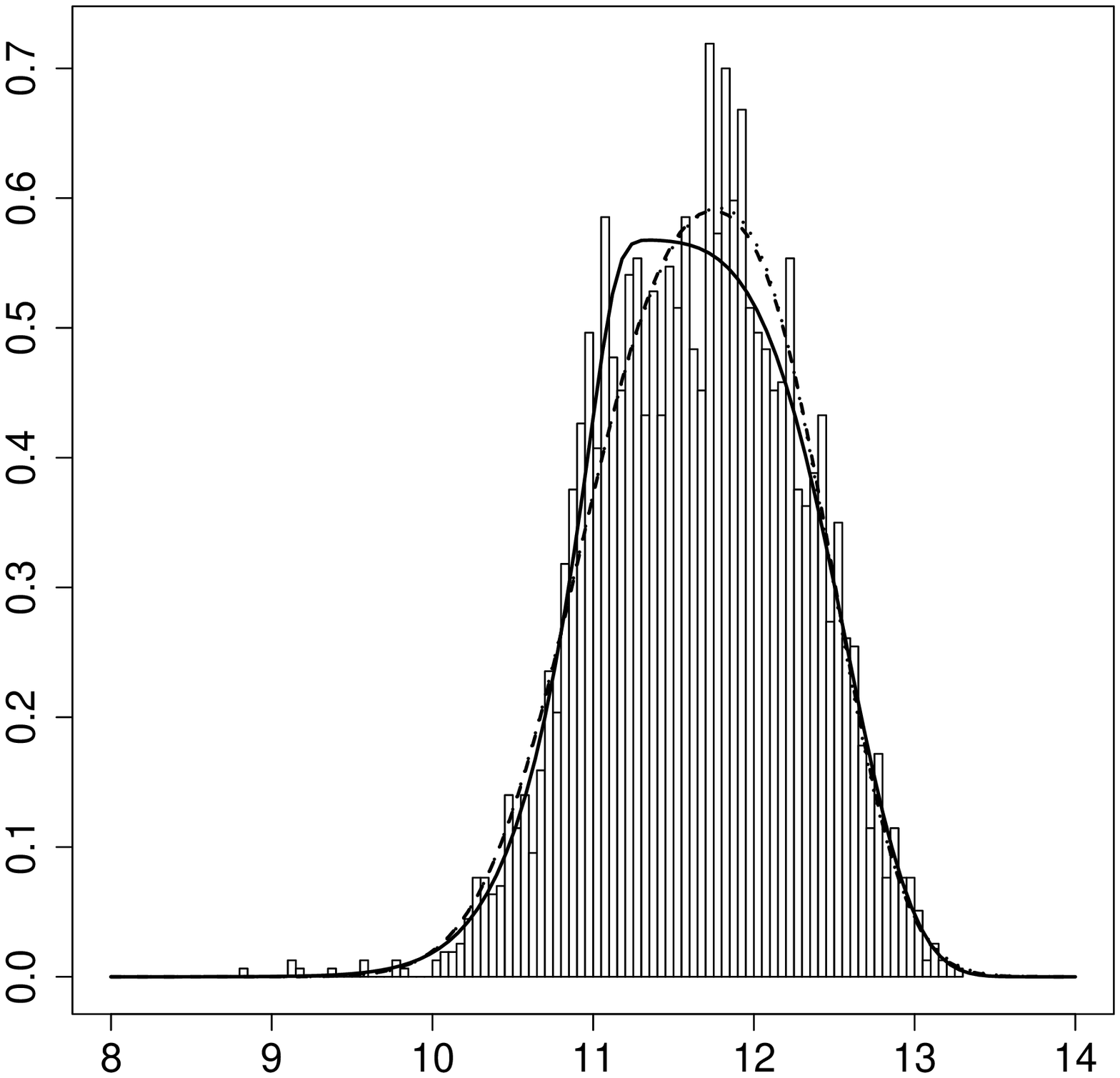,  height=5cm} &
\psfig{figure=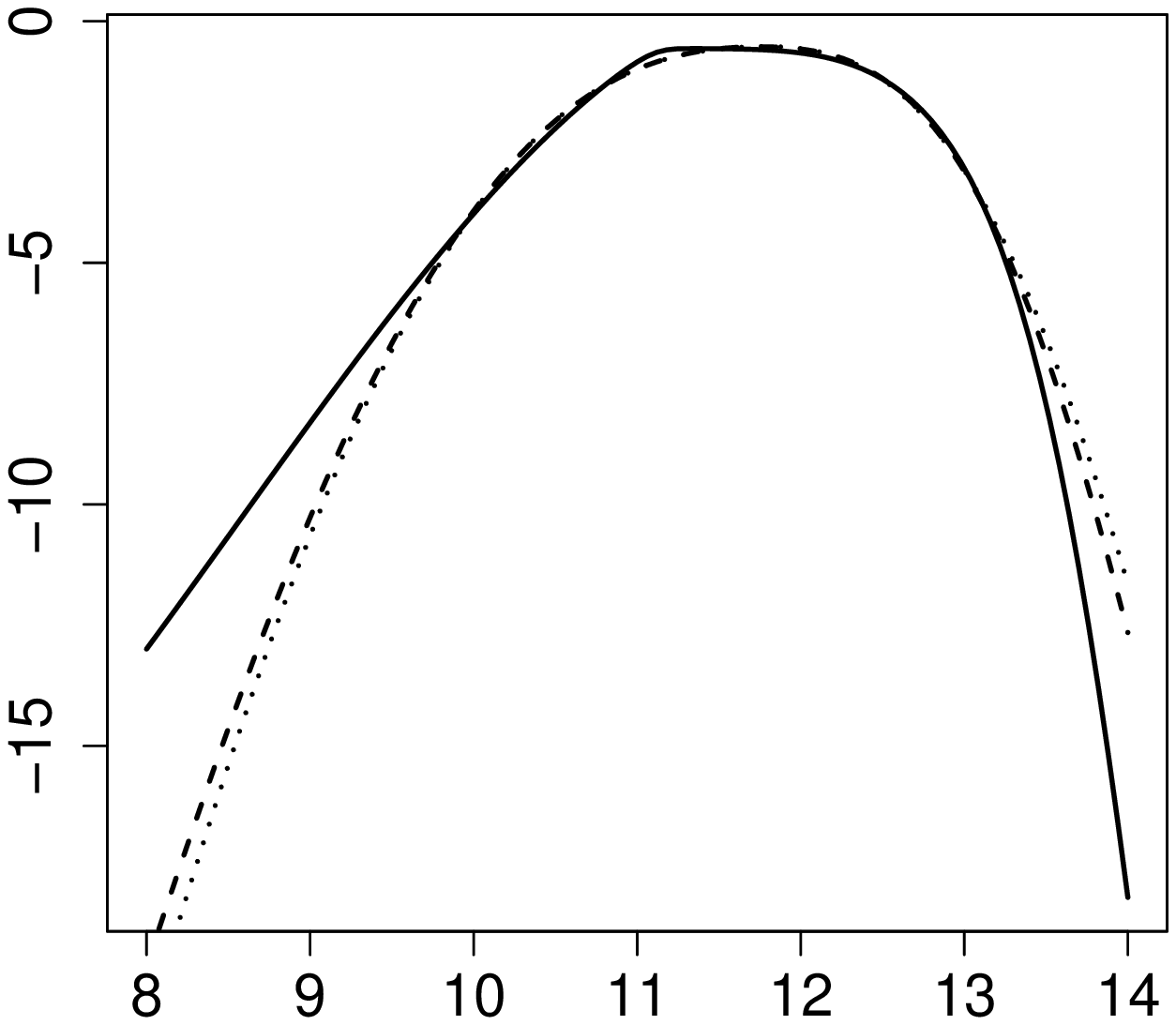,  height=5cm} \\
\end{tabular}
\end{center}
\caption{\small Internet traffic data (in logarithms; histogram) with (a) Predictive densities and (b) Log-predictive densities: DTP (continuous line); TPSH (dashed line); TPSC (dotted line).}
\label{fig:internetpred}
\end{figure}



\subsection{Actuarial Application}

In this application we analyse the claim sizes reported in \cite{B04} which can be found in \url{http:// lstat.kuleuven.be/Wiley/}. This data set contains $n=1823$ observations provided by the reinsurance brokers Aon Re Belgium. Such data typically contain extreme observations, and the logarithmic transformation is often used to reduce the effect of these extreme values \citep{R10}. A quantity of interest in this context is the probability that the claims exceed a certain bound \citep{V08}. This is often used for budgetary planning, which emphasises the importance of properly modelling the tails of the distribution. 

We explore two choices for $f$ in (\ref{dtppdf}): a Student-$t$ distribution and an SMN-BS distribution (see Appendix). We adopt the product prior structure (\ref{priorDTP2}) with uniform priors on $\gamma$ and $\zeta$. In order to produce matched priors on $\delta$ for these two models, we follow the strategy in Subsection \ref{PriorShapes}. The measure of kurtosis $\kappa\in(0.213,0.633)$ for the Student-$t$ model and $\kappa \in (0.213,0.560)$ for the SMN-BS model. Uniform priors for $\kappa$ induce compatible priors for $\delta$ in both models. Given that the data set contains a maximum number of $k=30$ repeated observations, we need to restrict the priors for $(\delta, \zeta)$: {for the Student-$t$ model we truncate $\delta>2$ and restrict $\zeta\in(-0.99,0.99)$. This truncation guarantees that condition (\ref{condt}) is satisfied since it implies that $\delta_1,\delta_2 > (k-1)/(n-k) \approx 0.02$.} For the SMN-BS model, the $\kappa$ measure is injective only on the interval $\delta\in(0,2.65)$, which covers the range $\kappa \in (0.213,0.560)$. In addition, for this model we can check that condition (\ref{condmix}) is satisfied if we truncate the $\delta_i$'s away from zero, {\it e.g.}~by imposing $\delta>1\times10^{-6}$ and taking $\zeta\in(-0.999,0.999)$.  Thus, we  restrict the prior for $\delta$ in the SMN-BS model to $(1\times10^{-6},2.65)$. The  posterior distributions are proper by Remark \ref{RemarkRepar}.


We also use the skew-$t$ distributions in \cite{AC03} ($s_{AC}$) and \cite{JF03}, denoted by $s_{JF}$ (see Appendix). Table \ref{table:AONMLE} shows the MLE and the AIC and BIC criteria, which favour the TPSC SMN-BS model overall. The Bayes factors, reported in Table \ref{table:BFAON}, favour the TPSC model for both underlying choices of $f$ and favours the  TPSC SMN-BS model overall, which agrees with the conclusion from AIC and BIC. However, there is no conclusive message from the SMN-BS models about which type of asymmetry is best for the data. The TPSH variant does almost as well. This is in line with the fact that the SMN-BS model does not distinguish clearly between TPSH and TPSC transformations, as discussed in Subsection 2.2. In contrast, the Student-$t$ models, for which both transformations are very distinct,  unambiguously indicate that the asymmetry is in the main body of the data and not in the tails: the TPSH $t$ model does very badly indeed, using both classical and Bayesian methods. Figure \ref{fig:AONpred} shows the corresponding predictive densities and illustrates the poor fit of the TPSH $t$ model which clearly affects the estimation of the right-tail probabilities shown in Figure \ref{fig:AONpred}(b): this model produces a predictive probability of 0.01 for the event $x>17$, while the other models lead to a predictive probability of less than 0.004. Unlike in the previous application, where right ``main-body'' skewness is combined with a heavier left tail (both $\gamma$ and $\zeta$ are estimated to be highly negative), the skew-$t$ by \cite{AC03} does well here, as these data combine right skewness in the main body with a fatter right tail. This is a feature that the $s_{AC}$ imposes (for positive $\lambda$ with both asymmetries in the opposite direction for $\lambda<0$). It is important to point out that the DTP families are not restricted in this way, as evidenced by the superiority of the DTP model in the previous application.

\begin{table}[h!]
\begin{center}
\begin{tabular}[h]{|c|c|c|c|c|c|c|c|}
\hline
Model & $\hat\mu$ & $\hat\sigma$ & $\hat\gamma$ & $\hat\delta$ & $\hat\zeta$& AIC & BIC\\
\hline
DTP $t$&  7.93 & 1.61 & -0.57 & 13.33 & 0.26 & 7283.1 & 7310.7\\
\hline
TPSC $t$&  7.90 & 1.62 & -0.59 & 10.98 & -- & 7281.6 & 7303.6\\
\hline
TPSH $t$& 9.13 & 1.46 & -- & 9998.80 & 0.99 & 7434.4 & 7456.5\\
\hline
DTP SMN-BS& 7.96  & 2.38 & -0.48 & 0.46 & -0.23& 7280.6 & 7308.2 \\
\hline
TPSC SMN-BS& 7.90 & 2.36 & -0.58 & 0.51 & -- & {\bf 7279.4} &  {\bf 7301.5}\\
\hline
TPSH SMN-BS& 8.03 & 3.43 & -- & 0.31 & -0.83 &  7280.1 & 7302.2\\
\hline
$s_{JF}$ &  1.56 & 0.02 & -- & ($\hat{a}$) 1560.6 & ($\hat{b}$) 5.07 & 7302.1 & 7324.1\\
\hline
$s_{AC}$ &  7.17 & 2.84 & ($\hat{\lambda}$) 4.90 & 13.75  & --  & 7280.7 & 7302.7\\
\hline
\end{tabular}
\caption{\small Aon data: Maximum likelihood estimates, AIC and BIC (best values in bold).}
\label{table:AONMLE}
\end{center}
\end{table}


\begin{table}[!h]
\begin{center}
\begin{tabular}[h]{|c|c|c|c|}
\hline
Model & DTP & TPSH & TPSC  \\
\hline
Student-$t$ & 1 &  5.00$\times10^{-65}$ & 2.05  \\
\hline
SMN-BS & 4.50 & 1.61 & 9.02 \\
\hline
\end{tabular}
\caption{\small Aon data: Bayes factors with respect to the DTP-$t$ model.}
\label{table:BFAON}
\end{center}
\end{table}

\begin{figure}[h]
\begin{center}
\begin{tabular}{c c}
\psfig{figure=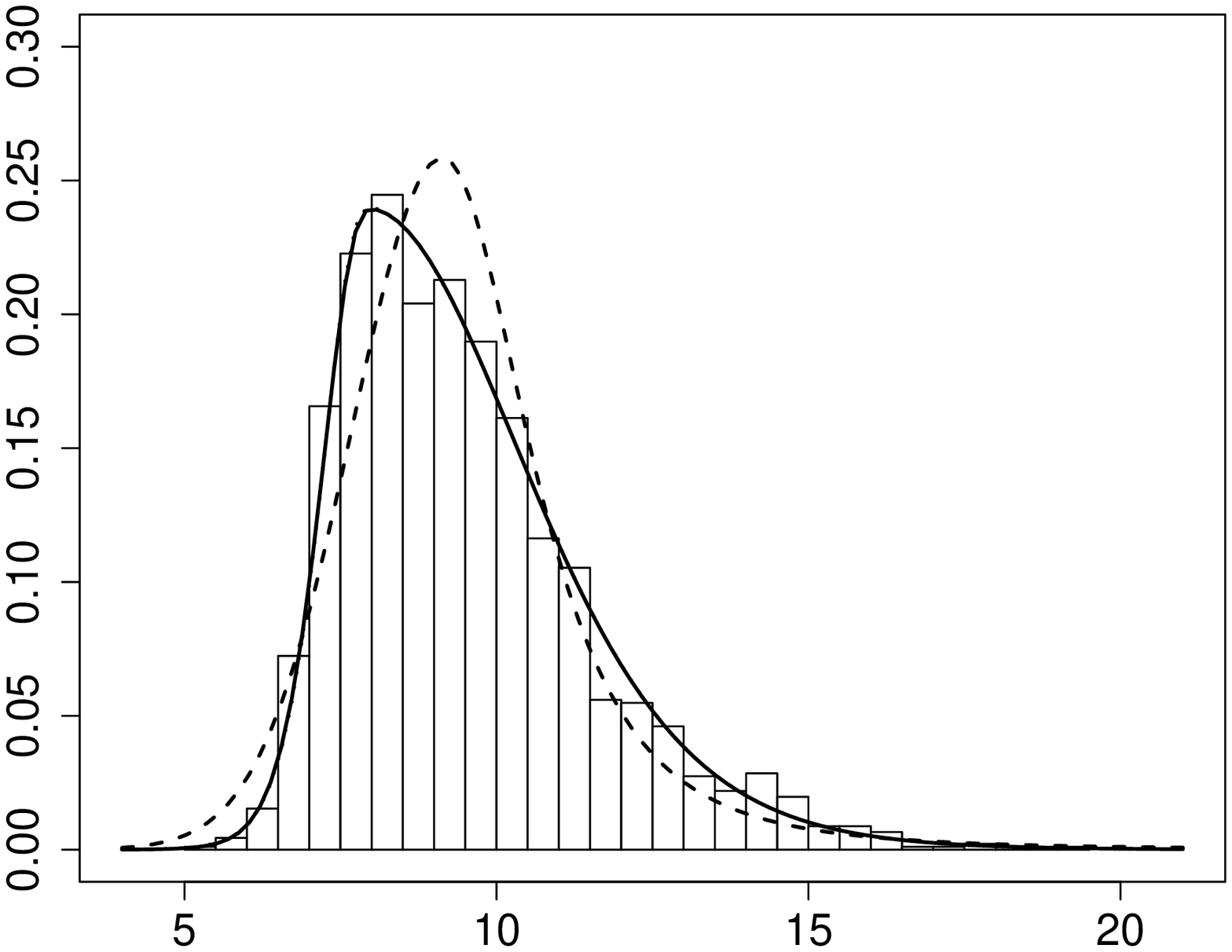,  height=5cm}  &
\psfig{figure=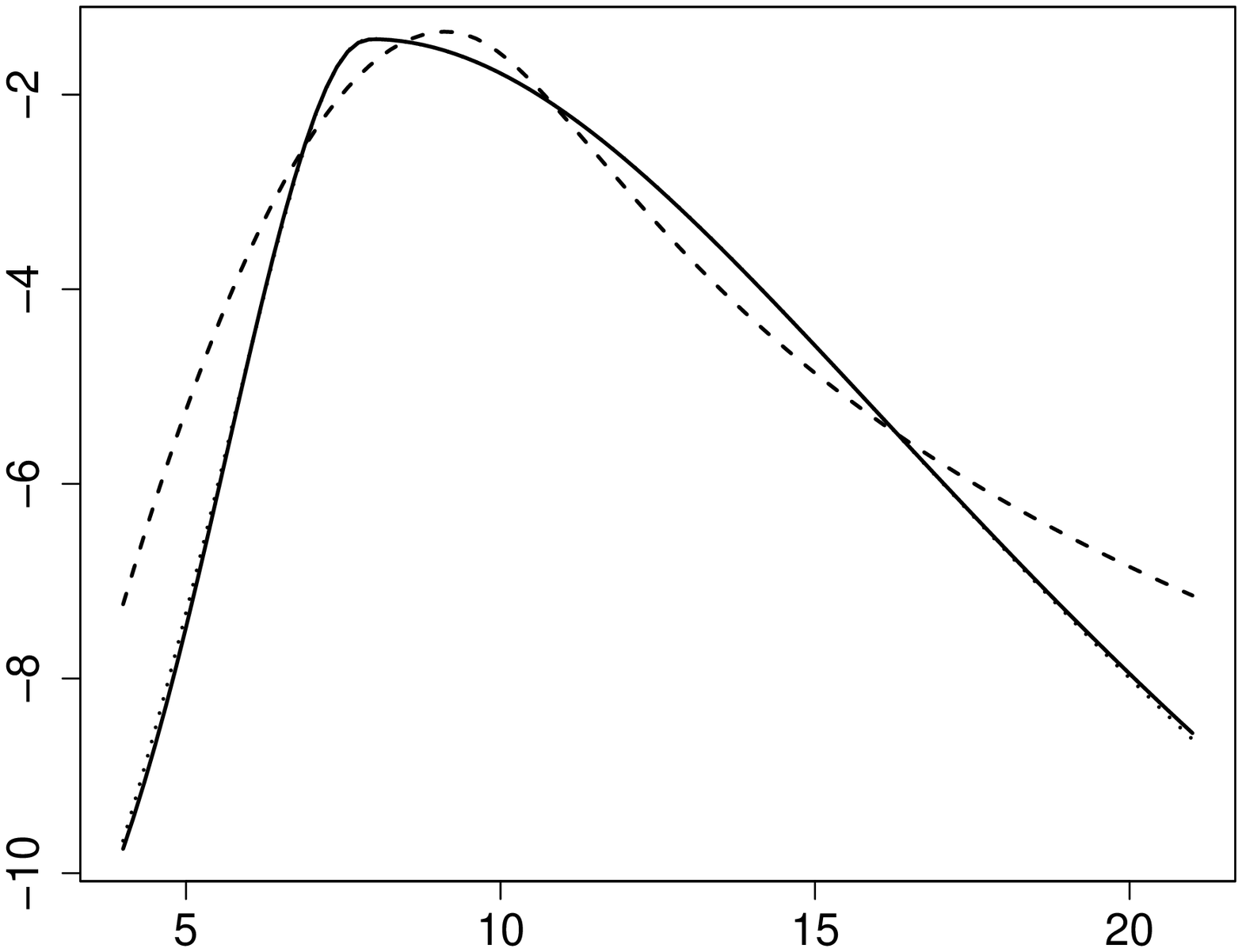,  height=5cm}\\
(a) & (b)\\
\psfig{figure=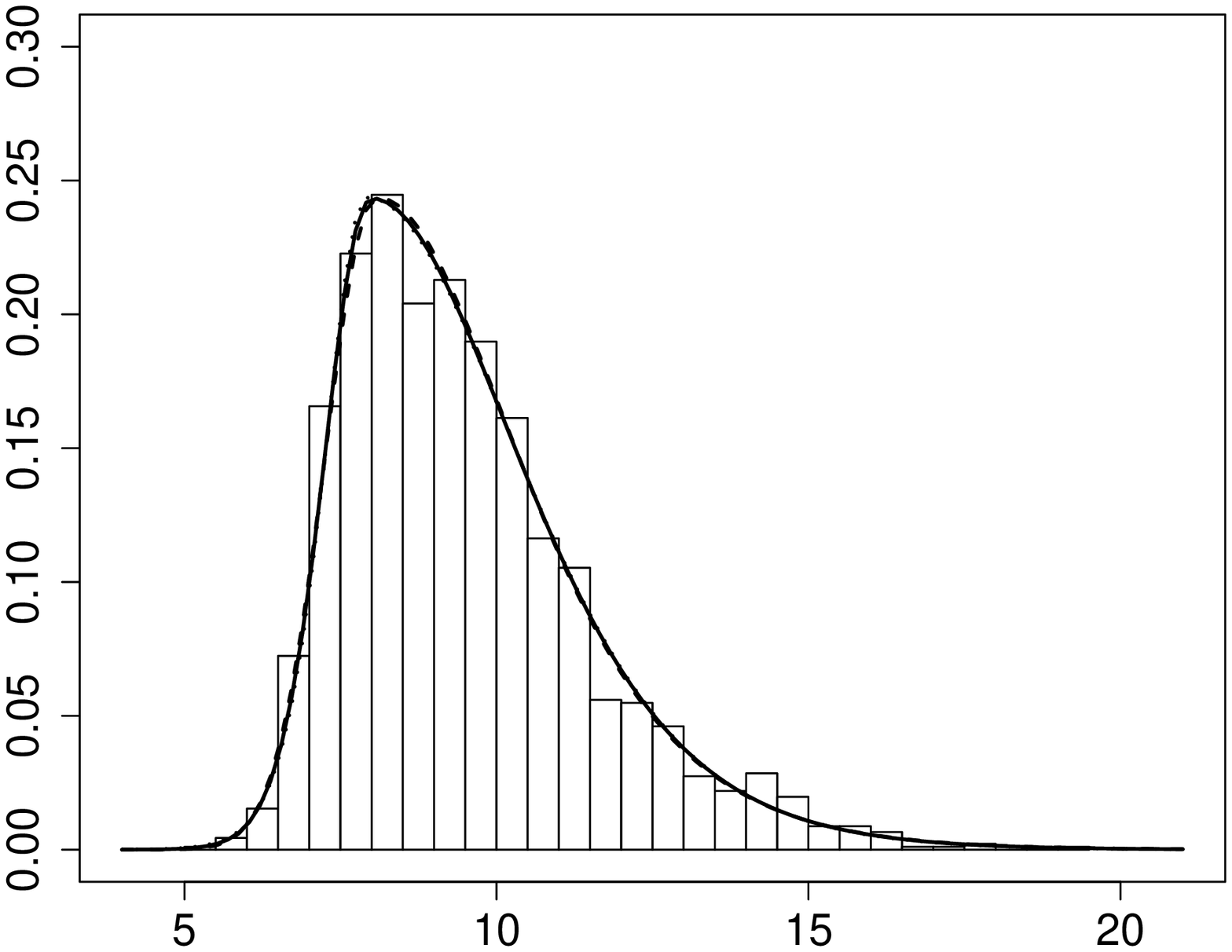,  height=5cm}  &
\psfig{figure=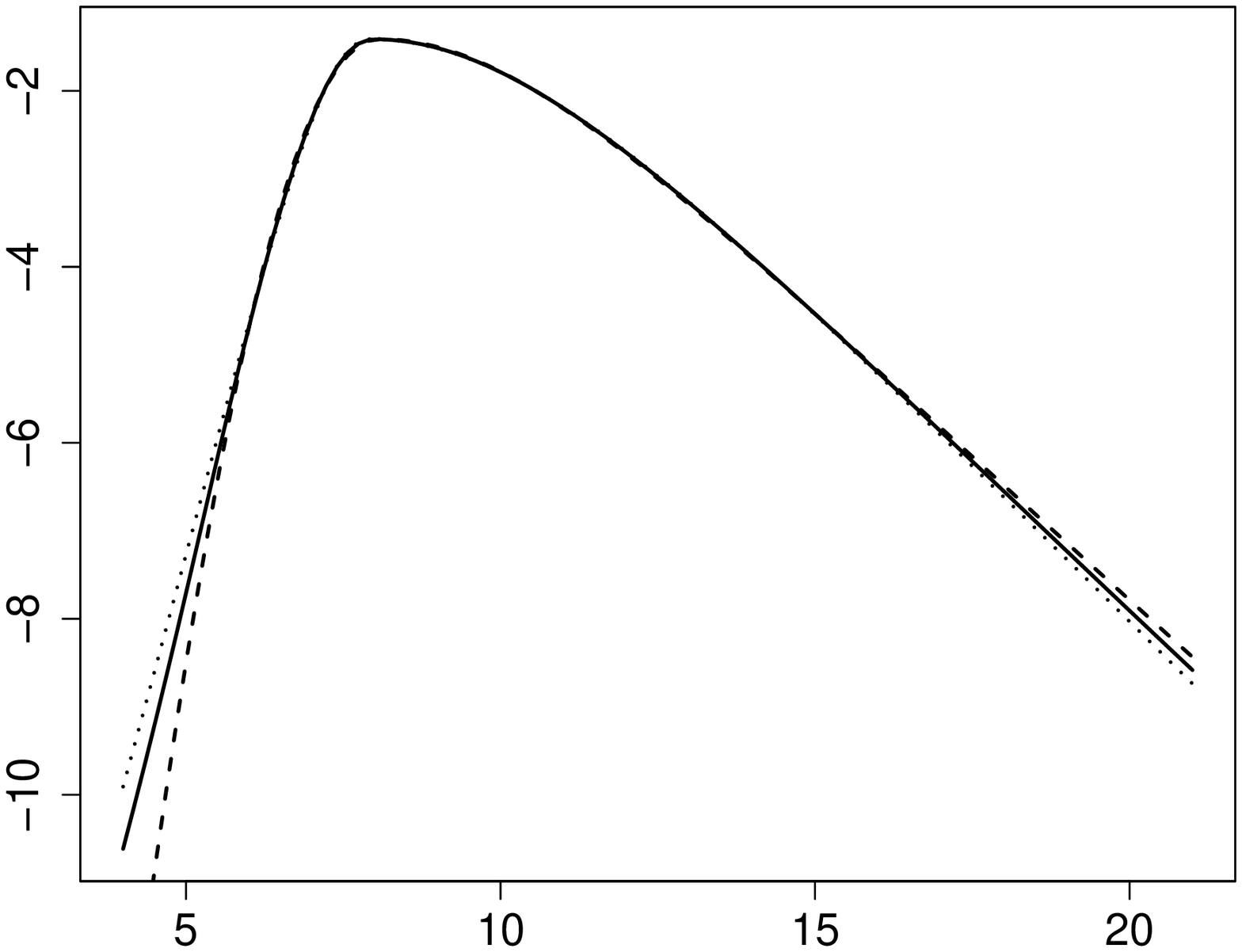,  height=5cm}\\
(c) & (d)
\end{tabular}
\end{center}
\caption{\small Aon data (histogram) with (a) Predictive densities and (b) Log-predictive densities: DTP $t$ (continuous line); TPSH $t$ (dashed line); TPSC $t$ (dotted line). (c) Predictive densities and (d) Log-predictive densities: DTP SMN-BS (continuous line); TPSH SMN-BS (dashed line); TPSC SMN-BS (dotted line).}
\label{fig:AONpred}
\end{figure}




\subsection{{Hierarchical Bayesian Models in Meta--Analysis}}

Bayesian hierarchical models are used in a variety of applied contexts to tackle parameter heterogeneity. A common example of this is the two--level normal model:
\begin{eqnarray}\label{oneway}
y_{j}\vert \theta_j &\sim& N(\theta_j,\sigma_j),\,\,\,j=1\dots n, \notag\\
\theta_j &\sim& N(\mu,\sigma).
\end{eqnarray}
A natural question is whether the assumption of normality of the random effects is appropriate: the implications of departures from this assumption are discussed in \cite{ZD01}, \cite{TL08} and \cite{MN11}.

In order to produce models that are robust to departures from normality of  $\theta_j$, several generalisations of (\ref{oneway}) have been proposed. For example, \cite{DH10} employ a Student--$t$ distribution, \cite{TL08} use a TPSC $t$ distribution with $\delta>2$ degrees of freedom, while \cite{D10} follows a Bayesian nonparametric approach. The use of non--normal distributional assumptions in this hierarchical model typically requires more sophisticated MCMC methods as discussed in \cite{RR09}.

\subsubsection{Fluoride Meta--analysis}

In this example we analyse the data set presented in \cite{M03} and used in \cite{TL08}, which contains $n=70$ trials assessing the effectiveness of fluoride toothpaste compared to a placebo conducted between 1954 and 1994. The treatment effect is the ``prevented fraction'', defined as the mean increment in the controls minus the mean increment in the treated group, divided by the mean increment in the controls. \cite{TL08} then propose the model
\begin{eqnarray}\label{hierarchical}
y_{j}\vert \theta_j &\sim& N(\theta_j,\sigma_j), \notag\\
\theta_j &\sim& P,
\end{eqnarray}
where $y_j$ is the estimate of the treatment effect in study $j$, $\theta_j$ is the true treatment effect in study $j$, and the parameters $\sigma_j$ are estimated from the data and assumed known. They compare the conclusions obtained for the true treatment effect for the following choices for $P$: (i) a TPSC $t$ distribution with $\delta>2$ degrees of freedom, (ii) a symmetric Student $t$ distribution with $\delta>2$ degrees of freedom, (iii) a TPSC normal distribution, and (iv) a normal distribution.

Here, we study six choices for  $P$: (i) a normal distribution, (ii) a symmetric sinh--arcsinh (SAS) distribution, (iii) a TPSC normal distribution, (iv) a TPSC SAS distribution \citep{R15}, (v) a TPSH SAS distribution and (vi) a DTP SAS distribution.
For the DTP model, we adopt the prior structure as in Subsection 3.3 $p(\mu,\sigma,\gamma,\delta,\zeta)=p(\mu)p(\sigma)p(\gamma)p(\delta)p(\zeta)$ with
$\mu \sim \text{Unif}(-10,10),
\sigma \sim \text{HalfCauchy}(0,s),
\gamma \sim \text{Unif}(-1,1),
\delta \sim p(\delta),
\zeta \sim \text{Unif}(-1,1)$,
with the prior shown in Figure \ref{fig:TSASPriors} for $\delta$, and $s=1/5,1,5$. For the simpler submodels we apply the same choices for the corresponding marginal priors. The results were not sensitive to the choice of $s$.

Figure \ref{fig:PredictiveDen} shows the posterior predictive densities for the treatment effect under different distributional assumptions for  the random effects. Clearly, symmetric distributions put more predictive  mass in the left tail $(-\infty,0.05)$ than those with asymmetry. Therefore, the probability of a small or a negative effect is overestimated under symmetric random effects. The predictive distributions obtained for DTP, TPSC, and TPSH SAS models are fairly similar in this case. However, the Bayes factors, shown in Table \ref{table:BFSDFluor}, slightly favour the TPSH SAS model, closely followed by the DTP SAS and the TPSC SAS models. Although the Bayes factors on the basis of this relatively small sample do not provide conclusive evidence about the best flexible model for the random effects, they definitely support asymmetric models with non-normal tails.


\begin{figure}[h]
\begin{center}
\begin{tabular}{c c c }
\psfig{figure=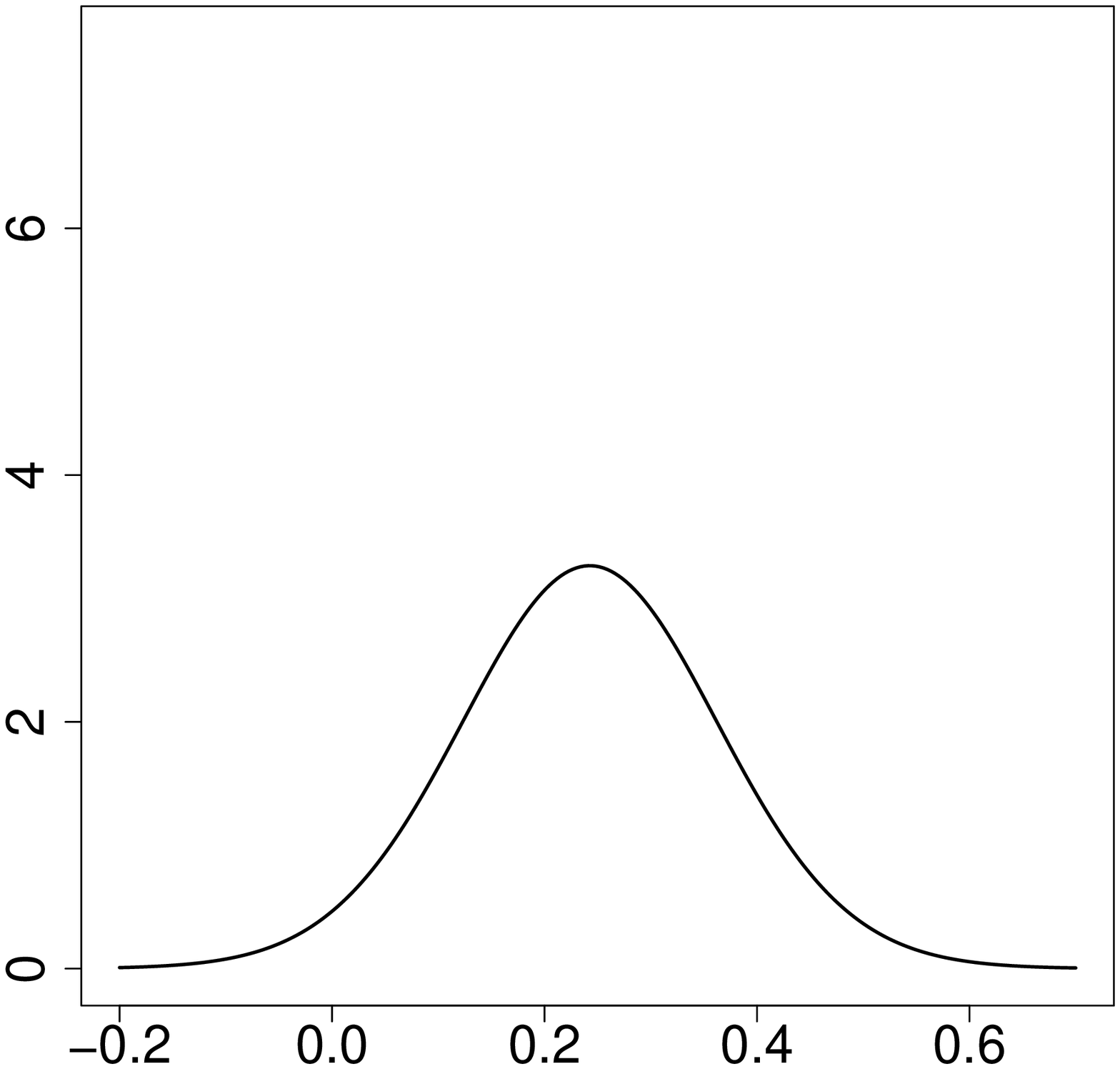,  height=4cm}  &
\psfig{figure=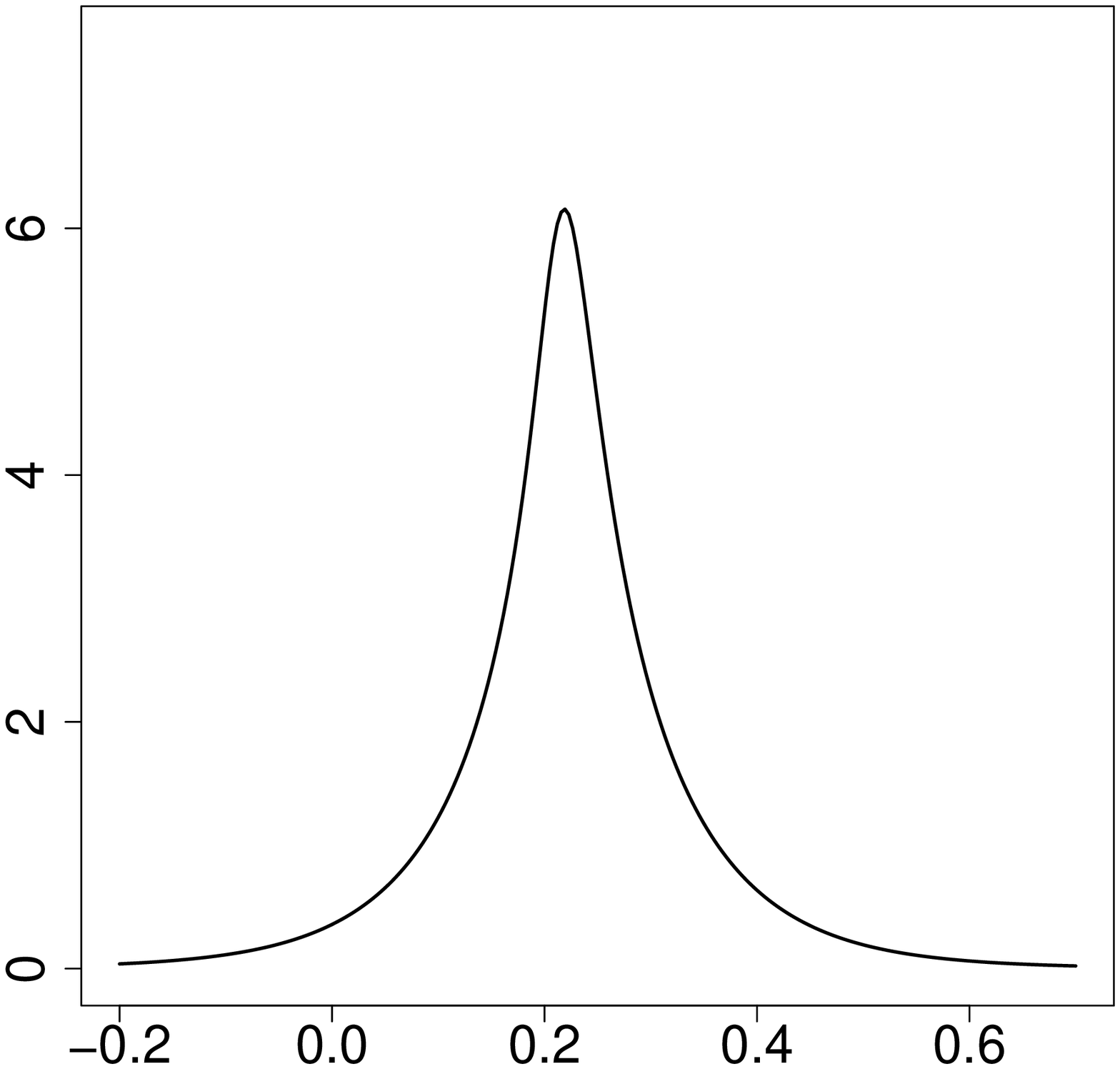,  height=4cm}  &
\psfig{figure=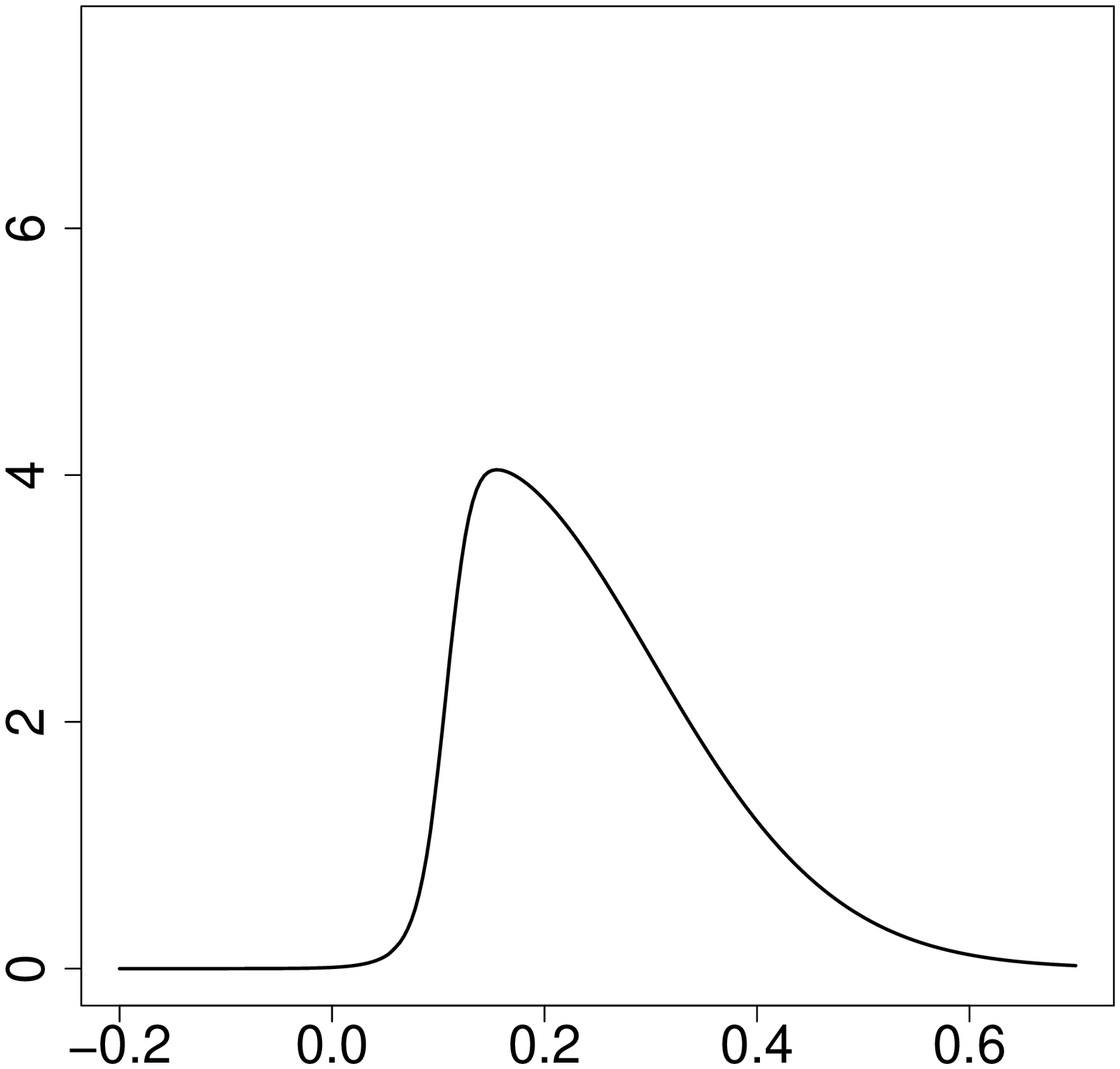,  height=4cm} \\
(a) & (b) & (c)\\
\psfig{figure=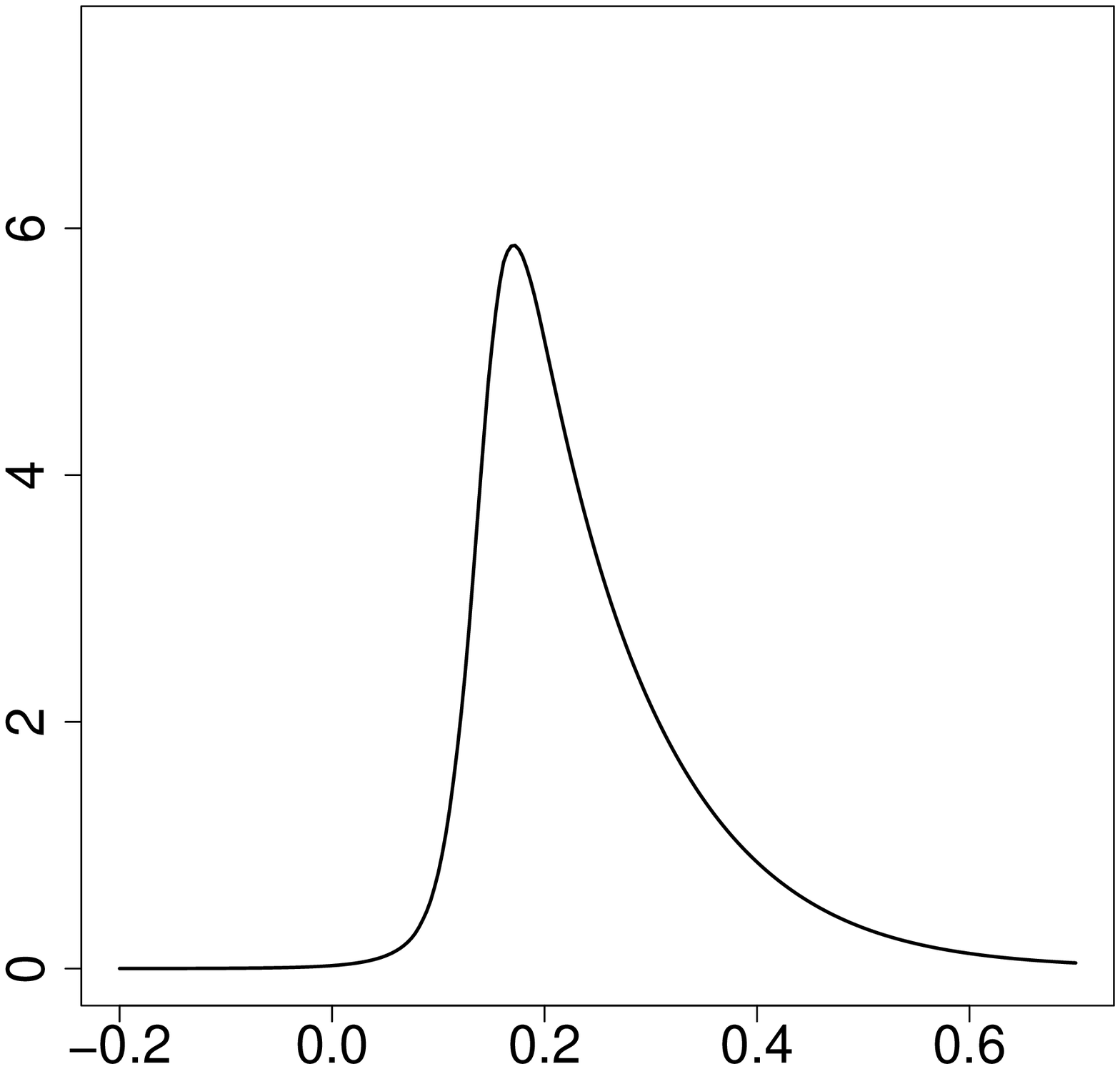,  height=4cm} &
\psfig{figure=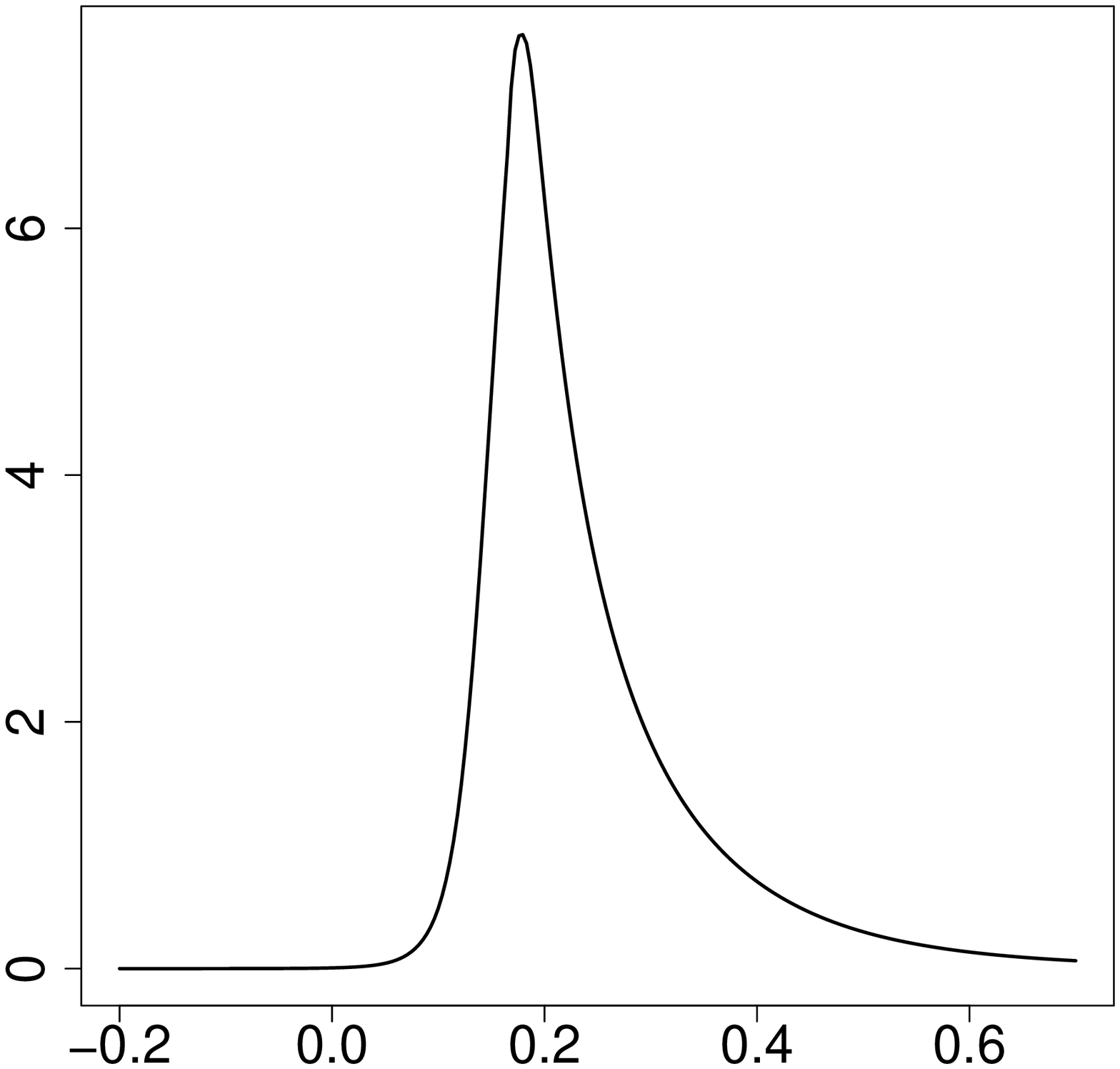, height=4cm} &
\psfig{figure=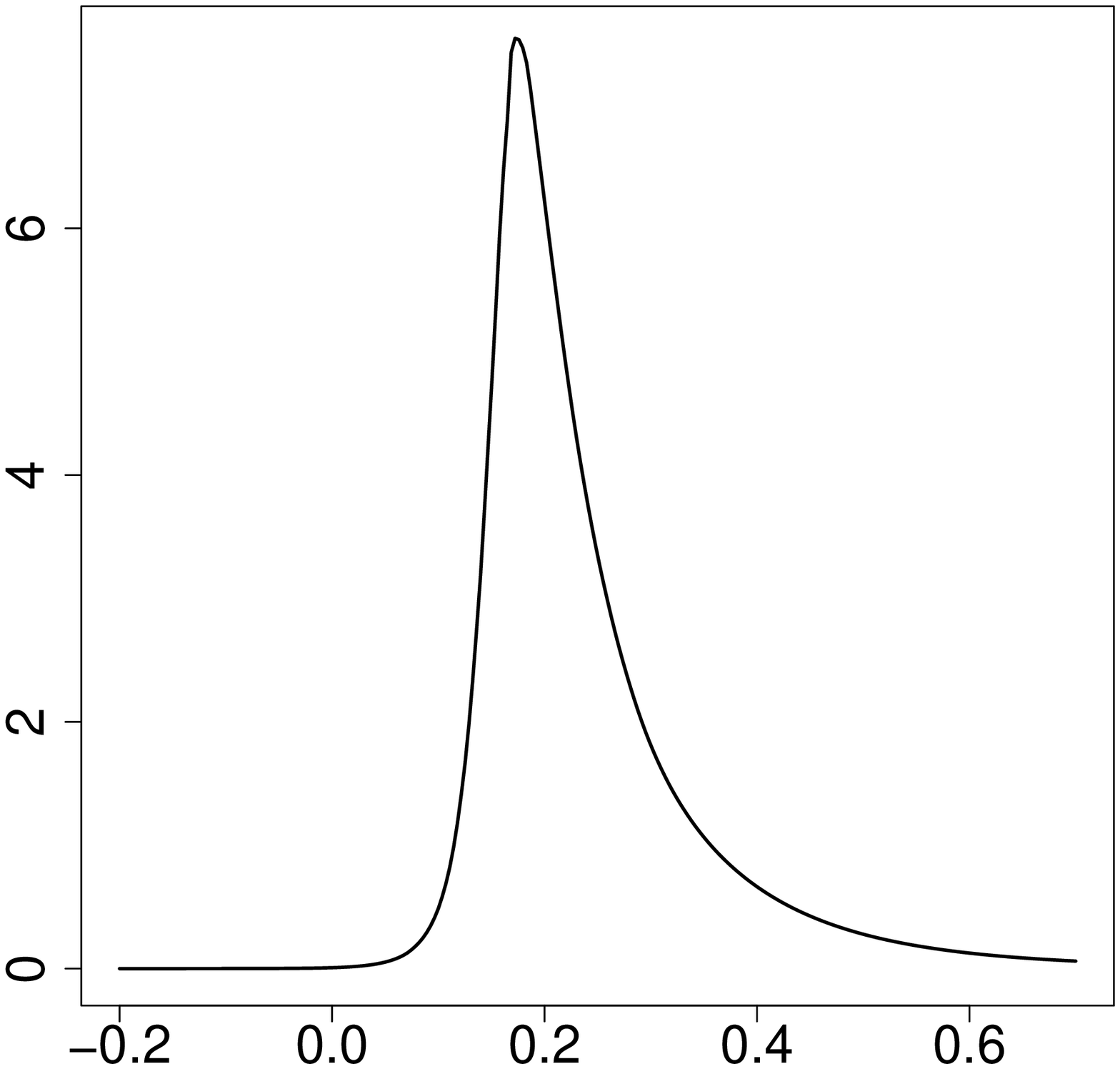,  height=4cm}\\
(d) & (e) & (f)\\
\end{tabular}
\end{center}
\caption{\small Predictive densities for the treatment effect: (a) Normal; (b) Symmetric SAS; (c) TPSC normal; (d) TPSC SAS; (e) TPSH SAS; (f) DTP SAS.}
\label{fig:PredictiveDen}
\end{figure}

\begin{table}[!h]
\begin{center}
\begin{tabular}[h]{|c|c|c|c|c|c|c|}
\hline
Model & DTP SAS & TPSH SAS & TPSC SAS & TPSC normal & Sym. SAS & normal \\
\hline
BF  & 1 &  1.27 & 0.30  & 0.05 & 0.02  &  5.2$\times 10^{-5}$ \\
\hline
\end{tabular}
\caption{\small Fluoride data: Bayes factors of submodels vs. the DTP SAS model.}
\label{table:BFSDFluor}
\end{center}
\end{table}

\section{Concluding Remarks}

We discuss a simple, intuitive and general class of transformations (DTP) that produces flexible unimodal and continuous distributions with parameters that separately control main-body skewness and tails on each side of the mode. Although some particular cases of DTP models have already appeared \citep{ZZ09,ZG10}, we formalise the idea and extend it to a wide range of symmetric ``base'' distributions ${\mathcal F}$. We also distinguish two subclasses of transformations and examine their interpretation as skewing mechanisms. A considerable advantage of the DTP class of transformations is the interpretability of its parameters (see \citealp{J14b} for the importance of interpretability) which, in the Bayesian context, also facilitates  prior elicitation. We propose a scale and location-invariant prior structure and derive conditions for posterior existence, also taking into account repeated and set observations. 

As illustrated by the applications, DTP families provide a flexible way of modelling unimodal data (or latent effects with unimodal distributions) and we provide a Bayesian framework for inference with sensible prior assumptions. In addition, we can conduct formal model comparison through Bayes factors for selecting models within the following classes:
\begin{itemize}
\item subclasses of DTP models with the same underlying symmetric base distribution $f$:  this is possible through the clearly separated roles of the parameters and the ensuing product prior structure with proper priors on $\gamma$ and $\zeta$. 
\item classes of DTP models with different underlying $f$: in nested cases this is easy, given the separate roles of the parameters and the ensuing product prior structure with proper priors on $\delta$, and in non-nested cases the priors on different shape parameters $\delta$ are matched through a common prior on the kurtosis measure $\kappa$.
\end{itemize}

DTP, TPSC and TPSH transformations can be used to construct robust models and, since they capture different kinds of asymmetry, selecting between these models provides more insight into the features of a data set. We have used Bayes factors for model choice, but other criteria, such as log-predictive scores, might be considered as well. \cite{ZZ09,ZG10} studied asymptotic properties of the maximum likelihood estimators (MLEs) for particular members of the DTP family (under the assumption of compactness of the parameter space). A more general study of the asymptotic properties of MLEs in DTP models represents an interesting research line.

DTP families can be extended to the multivariate case in several ways using general approaches. For TPSC models, \cite{FS07} propose the use of affine transformations to produce a multivariate extension while \cite{RS13a} propose to use copulas. In a similar fashion, the DTP (and consequently the TPSH) family can be used to construct multivariate distributions.

A different subclass of DTP transformations can be obtained by fixing $\sigma_1=\sigma$ and $\sigma_2=\dfrac{f(0;\delta_2)}{f(0;\delta_1)}\sigma$, leading to distributions with different shapes but equal mass cumulated on each side of the mode. This idea is proposed in \cite{R13}, who also composes this transformation with other skewing mechanisms to produce a different type of generalised skew-$t$ distribution.

\cite{RS13b} explore the use of Jeffreys priors in TPSC models. The use of Jeffreys priors for TPSH and DTP models is the object of further research. 

\section*{Acknowledgements}
We thank the Editor, an Associate Editor, and three referees for very helpful comments. We gratefully acknowledge research support from EPSRC grant EP/K007521/1.


\section*{Appendix}

\subsection*{Some density functions}
Throughout we use the notation $t=\dfrac{x-\mu}{\sigma}$.

\begin{enumerate}[(i)]
\item The symmetric Johnson-$\operatorname{S_U}$ distribution \citep{J49}:
\begin{eqnarray*}
\tilde f(x;\mu,\sigma,\delta) = \dfrac{\delta}{\sigma}\phi\left[\delta \operatorname{arcsinh}\left(t\right)\right]\left(1+t^2\right)^{-\frac{1}{2}}.
\end{eqnarray*}

\item The sinh-arcsinh distribution \citep{JP09}:
\begin{eqnarray*}
s_{JP}(x;\mu,\sigma,\delta) = \dfrac{\delta}{\sigma}\phi\left[\sinh\left(\delta \operatorname{arcsinh}\left(t\right)-\varepsilon\right)\right] \dfrac{\cosh\left(\delta \operatorname{arcsinh}\left(t\right)-\varepsilon\right)}{\sqrt{1+t^2}},
\end{eqnarray*}
where $\varepsilon\in{\mathbb R}$ controls the asymmetry of the density and symmetry corresponds to $\varepsilon=0$.




\item SMN-BS, a scale mixture of normals with Birnbaum-Saunders$(\delta,\delta)$ mixing:
\begin{eqnarray*}
\tilde f(x;\mu,\sigma,\delta) = \frac{e^{\frac{1}{\delta ^2}} \left(\sqrt{\delta  t^2+1}
   K_0\left(\frac{\sqrt{\delta  t^2+1}}{\delta
   ^2}\right)+K_1\left(\frac{\sqrt{\delta  t^2+1}}{\delta
   ^2}\right)\right)}{2 \pi \sigma  \delta ^{3/2} \sqrt{\delta  t^2+1}}.
\end{eqnarray*}
where $K_n(z)$ represents the modified Bessel function of the second kind.

\item  The skew-$t$ density from \cite{JF03}:
\begin{eqnarray*}
s_{JF}(x;\mu,\sigma,a,b)=C_{a,b}^{-1}\left[1+\dfrac{t}{\sqrt{a+b+t^2}}\right]^{a+1/2}\left[1-\dfrac{t}{\sqrt{a+b+t^2}}\right]^{b+1/2},
\end{eqnarray*}
where $a,b>0$, and $C_{a,b}=2^{a+b-1}\operatorname{Beta}(a,b)\sqrt{a+b}$. The parameters $(a,b)$ control the tails and skewness jointly. The density $s_{JF}$ is asymmetric if and only if $a\neq b$, so that the density is skewed only when the tail behaviour differs in each direction. 

\item  The skew-$t$ density from \cite{AC03}:
\begin{eqnarray*}
s_{AC}(x;\mu,\sigma,\lambda,\delta)= 2f(x;\mu,\sigma,\delta)F\left(\lambda x \sqrt{\dfrac{\delta+1}{\delta+x^2}};\mu,\sigma,\delta+1\right),
\end{eqnarray*}
where $\lambda \in{\mathbb R}$ and $f$ and $F$ are, respectively, the Student-$t$ density function and the Student-$t$ distribution function.

\end{enumerate}

\subsection*{Proofs}

In the proofs below, equation numbers other than (\ref{LowerBound}) refer to equations in the main paper.

\subsubsection*{Proof of Theorem 1}

The marginal likelihood of the data can be  bounded from below as follows

\begin{eqnarray}\label{LowerBound}
m({\bf x}) &\propto& \int_{\Delta}\int_{\Delta}\int_{\Gamma}\int_{{\mathbb R}_+}\int_{{\mathbb R}} \left[\prod_{j=1}^n s(x_j;\mu,\sigma,\gamma,\delta_1,\delta_2) \right] p(\mu)p(\sigma)p(\gamma)p(\delta_1)p(\delta_2)\, d\mu d\sigma d\gamma d\delta_1 d\delta_2 \notag\\
&\geq& \int_{\Delta}\int_{\Delta}\int_{\Gamma}\int_{{\mathbb R}_+}\int_{-\infty}^{x_{(1)}} \dfrac{f(0;\delta_1)^n}{\sigma^n H(\gamma)^n\left[f(0;\delta_1)+f(0;\delta_2)\right]^n} \left[\prod_{j=1}^n f\left(\dfrac{x_j-\mu}{\sigma a(\gamma)};\delta_2\right) \right] \notag\\
&\times&p(\mu)p(\sigma)p(\gamma)p(\delta_1)p(\delta_2)\, d\mu d\sigma d\gamma d\delta_1 d\delta_2
\end{eqnarray}

\noindent where $s(\cdot)$ is given by (8) in the paper, $H(\gamma)=\max\{a(\gamma),b(\gamma)\}$, and $x_{(1)}$ represents the smallest order statistic of ${\bf x}$. Therefore:

\begin{enumerate}[(i)]
\item follows by noting that the lower bound (\ref{LowerBound}) does not depend upon $\delta_1$.

\item follows by using the following inequality, provided $f(0;\delta)\le U$ for some $U>0$
\begin{eqnarray*}
\dfrac{f(0;\delta_1)^n}{\left[f(0;\delta_1)+f(0;\delta_2)\right]^n} \geq \dfrac{f(0;\delta_1)^n}{2^n U^n},
\end{eqnarray*}
which leads to the necessary condition (11).

\item Given that $f(0;\delta)$ is continuous and monotonic, then for any $0\leq \inf_{\delta\in\Delta}f(0;\delta)<M < \sup_{\delta\in\Delta}f(0;\delta)$, there exists a set $\Delta_2(M)\subset \Delta$ such that $f(0;\delta)<M$ for all $\delta\in \Delta_2(M)$. If we integrate $\delta_2$ over $\Delta_2$, we obtain the following lower bound, up to a proportionality constant, for $m({\bf x})$

\begin{eqnarray*}
&&\int_{\Delta_2}\int_{\Delta}\int_{\Gamma}\int_{{\mathbb R}_+}\int_{-\infty}^{x_{(1)}} \dfrac{f(0;\delta_1)^n}{\sigma^n H(\gamma)^n\left[f(0;\delta_1)+f(0;\delta_2)\right]^n} \left[\prod_{j=1}^n f\left(\dfrac{x_j-\mu}{\sigma a(\gamma)};\delta_2\right) \right]\\
&\times&p(\mu)p(\sigma)p(\gamma)p(\delta_1)p(\delta_2)\, d\mu d\sigma d\gamma d\delta_1 d\delta_2\\
&\geq& \int_{\Delta_2}\int_{\Delta}\int_{\Gamma}\int_{{\mathbb R}_+}\int_{-\infty}^{x_{(1)}} \dfrac{f(0;\delta_1)^n}{\sigma^n H(\gamma)^n\left[f(0;\delta_1)+M\right]^n} \left[\prod_{j=1}^n f\left(\dfrac{x_j-\mu}{\sigma a(\gamma)};\delta_2\right) \right]\\
&\times&p(\mu)p(\sigma)p(\gamma)p(\delta_1)p(\delta_2)\, d\mu d\sigma d\gamma d\delta_1 d\delta_2.
\end{eqnarray*}

From the last expression we obtain the necessary condition (12).
\end{enumerate}

Analogous results can be obtained for $\delta_2$ by integrating $\mu$ over $(x_{(n)},\infty)$, where $x_{(n)}$ represents the largest order statistic of ${\bf x}$.

\subsubsection*{Proof of Theorem 2}


\begin{enumerate}[(i)]

\item In this parameterization, $\varepsilon$ in (2) does not depend on $\sigma$. This fact will be used implicitly in a change of variable below.
We obtain
\begin{eqnarray*}
p({\bf x}) &\propto& \int_{\Delta} \int_{\Delta} \int_0^{\infty} \int_{-\infty}^{\infty} \int_{{\mathbb R}^n_+} \dfrac{1}{[a(\gamma)+b(\gamma)]^n}\dfrac{\prod_{j=1}^n\lambda_j^{\frac{1}{2}}}{\sigma^{n+1}}\exp\left[-\dfrac{1}{2\sigma^2}\sum_{j=1}^n \dfrac{\lambda_j}{i_j(\gamma)^2}(x_j-\mu)^2\right] \\
&\times& p(\gamma\delta_1,\delta_2)\prod_{j=1}^n\left\{\varepsilon dP_{\lambda_j\vert\delta_1} I(x_j<\mu) + (1-\varepsilon) dP_{\lambda_j\vert\delta_2} I(x_j\geq\mu) \right\}d\mu d\sigma d\gamma d\delta_1 d\delta_2\\
&\leq& \int_{\Delta} \int_{\Delta} \int_0^{\infty} \int_{-\infty}^{\infty} \int_{{\mathbb R}^n_+} \dfrac{1}{[a(\gamma)+b(\gamma)]^n}\dfrac{\prod_{j=1}^n\lambda_j^{\frac{1}{2}}}{\sigma^{n+1}}\exp\left[-\dfrac{1}{2\sigma^2h(\gamma)^2}\sum_{j=1}^n \lambda_j(x_j-\mu)^2\right] \\
&\times& p(\gamma)p(\delta_1,\delta_2)\prod_{j=1}^n\left\{\varepsilon dP_{\lambda_j\vert\delta_1} I(x_j<\mu) + (1-\varepsilon) dP_{\lambda_j\vert\delta_2} I(x_j\geq\mu) \right\}d\mu d\sigma d\gamma d\delta_1 d\delta_2,
\end{eqnarray*}
\noindent where $i_j(\gamma)= a(\gamma)I(x_j\geq \mu) + b(\gamma)I(x_j<\mu)$ and $h(\gamma)=\max\{a(\gamma),b(\gamma)\}$. Now, consider the change of variable $\theta=\sigma h(\gamma)$, then we get that this upper bound can be written as follows
\begin{eqnarray*}
&&\int_{\Delta} \int_{\Delta} \int_0^{\infty} \int_{-\infty}^{\infty} \int_{{\mathbb R}^n_+} \dfrac{h(\gamma)^n}{[a(\gamma)+b(\gamma)]^n}\dfrac{\prod_{j=1}^n\lambda_j^{\frac{1}{2}}}{\theta^{n+1}}\exp\left[-\dfrac{1}{2\theta^2}\sum_{j=1}^n \lambda_j(x_j-\mu)^2\right] \\
&\times& p(\gamma)p(\delta_1,\delta_2) \prod_{j=1}^n\left\{\varepsilon dP_{\lambda_j\vert\delta_1} I(x_j<\mu) + (1-\varepsilon) dP_{\lambda_j\vert\delta_2} I(x_j\geq\mu) \right\}d\mu d\theta d\gamma d\delta_1 d\delta_2.
\end{eqnarray*}
By using that $0\leq \varepsilon\leq 1$, $\dfrac{1}{2}\leq \dfrac{h(\gamma)^n}{[a(\gamma)+b(\gamma)]^n} \leq 1$ it follows that the propriety of the posterior of $(\mu,\sigma,\gamma,\delta_1,\delta_2)$ under this prior structure is equivalent to the propriety of the posterior distribution of a TPSH sampling model with parameters $(\mu,\sigma,\delta_1,\delta_1)$ and prior structure $\pi(\mu,\sigma,\delta_1,\delta_1)\propto\sigma^{-1}p(\delta_1,\delta_2)$, where $p(\delta_1,\delta_2)$ is a proper prior. The rest of the proof thus focuses on the latter model, for which, by construction, we have
\begin{eqnarray*}
f(x_j; \mu,\sigma,\delta_1,\delta_2) &=& \int_0^{\infty} \dfrac{2\lambda_j^{\frac{1}{2}}}{\sqrt{2\pi}\sigma} \exp \left[-\dfrac{\lambda_j}{2\sigma^2}(x_j-\mu)^2\right]\\
&\times&\,\left\{\varepsilon dP_{\lambda_j\vert\delta_1} I(x_j<\mu) + (1-\varepsilon) dP_{\lambda_j\vert\delta_2} I(x_j\geq\mu) \right\},
\end{eqnarray*}
\noindent with $\varepsilon$ as in (2). Then, we can write the marginal of ${\bf x}$ as follows
\begin{eqnarray*}
p({\bf x}) &\propto& \int_{\Delta} \int_{\Delta} \int_0^{\infty} \int_{-\infty}^{\infty} \int_{{\mathbb R}^n_+} \dfrac{\prod_{j=1}^n\lambda_j^{\frac{1}{2}}}{\sigma^{n+1}}\exp\left[-\dfrac{1}{2\sigma^2}\sum_{j=1}^n \lambda_j(x_j-\mu)^2\right]p(\delta_1,\delta_2)  \\
&\times&\prod_{j=1}^n\left\{\varepsilon dP_{\lambda_j\vert\delta_1} I(x_j<\mu) + (1-\varepsilon) dP_{\lambda_j\vert\delta_2} I(x_j\geq\mu) \right\}d\mu d\sigma d\delta_1 d\delta_2.
\end{eqnarray*}
Separating the integral with respect to $\mu$ into $n+1$ integrals over the domains $(-\infty,x_{(1)})$, $[x_{(1)},x_{(2)})$, ..., $[x_{(n)},\infty)$, we have that
\begin{eqnarray*}
I_1 &=& \int_{\Delta} \int_{\Delta} \int_0^{\infty} \int_{-\infty}^{x_{(1)}} \int_{{\mathbb R}^n_+} \dfrac{\prod_{j=1}^n\lambda_j^{\frac{1}{2}}}{\sigma^{n+1}}\exp\left[-\dfrac{1}{2\sigma^2}\sum_{j=1}^n \lambda_j(x_j-\mu)^2\right]\\
&\times&p(\delta_1,\delta_2)(1-\varepsilon)^n \, \prod_{j=1}^n dP_{\lambda_j\vert\delta_2} d\mu d\sigma d\delta_1 d\delta_2.
\end{eqnarray*}
By noting that $0\leq\varepsilon\leq 1$, extending the integration domain on $\mu$ to the whole real line and integrating out $\delta_1$ we obtain
\begin{eqnarray*}
I_1\leq \int_{\Delta}\int_0^{\infty} \int_{-\infty}^{\infty}\int_{{\mathbb R}^n_+} \dfrac{\prod_{j=1}^n\lambda_j^{\frac{1}{2}}}{\sigma^{n+1}}\exp\left[-\dfrac{1}{2\sigma^2}\sum_{j=1}^n \lambda_j(x_j-\mu)^2\right] p(\delta_2) \, \prod_{j=1}^n dP_{\lambda_j\vert\delta_2} d\mu d\sigma d\delta_2<\infty.
\end{eqnarray*}
The finiteness of this integral is obtained using Theorem 1 from \cite{FS98b}. Now, using similar arguments we have that
\begin{eqnarray*}
I_2 &=& \int_{\Delta} \int_{\Delta} \int_0^{\infty} \int_{x_{(n)}}^{\infty} \int_{{\mathbb R}^n_+} \dfrac{\prod_{j=1}^n\lambda_j^{\frac{1}{2}}}{\sigma^{n+1}}\exp\left[-\dfrac{1}{2\sigma^2}\sum_{j=1}^n \lambda_j(x_j-\mu)^2\right]\\
&\times&p(\delta_1,\delta_2)\varepsilon^n \, \prod_{j=1}^n dP_{\lambda_j\vert\delta_1}d\mu d\sigma d\delta_1 d\delta_2\\
&\leq& \int_{\Delta} \int_0^{\infty} \int_{-\infty}^{\infty} \int_{{\mathbb R}^n_+} \dfrac{\prod_{j=1}^n\lambda_j^{\frac{1}{2}}}{\sigma^{n+1}}\exp\left[-\dfrac{1}{2\sigma^2}\sum_{j=1}^n \lambda_j(x_j-\mu)^2\right]\\
&\times&p(\delta_1) \, \prod_{j=1}^n dP_{\lambda_j\vert\delta_1}d\mu d\sigma d\delta_1  < \infty.
\end{eqnarray*}

Finally, for an intermediate region we have
\begin{eqnarray*}
I_3 &=& \int_{\Delta} \int_{\Delta}\int_0^{\infty} \int_{x_{(k)}}^{x_{(k+1)}}\int_{{\mathbb R}^n_+} \dfrac{\prod_{j=1}^n\lambda_j^{\frac{1}{2}}}{\sigma^{n+1}}\exp\left[-\dfrac{1}{2\sigma^2}\sum_{j=1}^n \lambda_j(x_{(j)}-\mu)^2\right]p(\delta_1,\delta_2)\varepsilon^k (1-\varepsilon)^{n-k}   \\
&\times& \prod_{j=1}^k dP_{\lambda_j\vert\delta_1} \prod_{j=k+1}^n dP_{\lambda_j\vert\delta_2}d\mu d\sigma d\delta_1 d\delta_2\\
&\leq& \int_{\Delta} \int_{\Delta} \int_0^{\infty} \int_{-\infty}^{\infty} \int_{{\mathbb R}^n_+} \dfrac{\prod_{j=1}^n\lambda_j^{\frac{1}{2}}}{\sigma^{n+1}}\exp\left[-\dfrac{1}{2\sigma^2}\sum_{j=1}^n \lambda_j(x_{(j)}-\mu)^2\right]p(\delta_1,\delta_2) \\
&\times& \prod_{j=1}^k dP_{\lambda_j\vert\delta_1} \prod_{j=k+1}^n dP_{\lambda_j\vert\delta_2}d\mu d\sigma d\delta_1 d\delta_2 < \infty.
\end{eqnarray*}
The finiteness follows again from Theorem 1 from \cite{FS98b}. Combining the finiteness of $I_1$, $I_2$ and $I_3$ the result follows.

\item This follows by using the previous proof together with Theorems 1, 2, and 3 from \cite{FS98b}.
\end{enumerate}

\subsubsection*{Proof of Corollary 1}

From the proof of point (i) in Theorem 2 it follows that the propriety of the posterior distribution of $(\mu,\sigma,\gamma,\delta_1,\delta_2)$ is equivalent to proving the propriety of $(\mu,\sigma,\delta)$, assuming that $S_1,\dots,S_n$ is an {i.i.d.}~sample of set observations from a scale mixture of normals $f(\cdot; \mu,\sigma,\delta)$ and adopting the prior $\pi(\mu,\sigma,\delta) \propto \sigma^{-1}p(\delta)$, where $p(\delta)$ is proper. The result then follows by combining this fact with Theorem 4 from \cite{FS98b}.



\end{document}